\documentclass[%
twocolumn,
superscriptaddress,
nobibnotes,
 amsmath,amssymb,
prb,
floatfix,
]{revtex4-2}

\usepackage{graphicx}% Include figure files
\usepackage{dcolumn}% Align table columns on decimal point
\usepackage{bm}% bold math
\usepackage{gensymb}
\usepackage{soul}
\usepackage{xcolor}
\usepackage{chemarr} %Double arrow with labels
\newcommand*\un[1]{\,\mathrm{#1}}

\def \be {\begin{equation}}
\def \ee {\end{equation}}
\def \bea {\begin{eqnarray}}
\def \eea {\end{eqnarray}}

\begin{document}

\preprint{APS/123-QED}

\title{Response of the chiral soliton lattice to spin polarized currents}

\author{S. A. Osorio}
\affiliation{ 
Instituto de Nanociencia y Nanotecnolog\'ia (CNEA-CONICET), Nodo Bariloche, Av. Bustillo 9500 (R8402AGP), S. C. de Bariloche, R\'io Negro, Argentina
% Instituto de Nanociencia y Nanotecnolog\'ia, CNEA-CONICET, Centro At\'omico Bariloche, (R8402AGP) S. C. de Bariloche, R\'io Negro, Argentina%\\This line break forced with \textbackslash\textbackslash
}
\affiliation{
Gerencia de F\'isica,  Centro Atómico Bariloche, Av. Bustillo 9500 (R8402AGP), S. C. de Bariloche, R\'io Negro, Argentina
}

\author{A. Athanasopoulos}%
\affiliation{ 
Aragon Nanoscience and Materials Institute (CSIC‐University of Zaragoza) and Condensed Matter Physics Department, University of Zaragoza, C/ Pedro Cerbuna 12, 50009 Zaragoza, Spain
}

\author{V. Laliena}
\affiliation{
Department of Applied Mathematics, University of Zaragoza, C/ María de Luna 3, 50018 Zaragoza, Spain
}%
\author{J. Campo}
\affiliation{ 
Aragon Nanoscience and Materials Institute (CSIC‐University of Zaragoza) and Condensed Matter Physics Department, University of Zaragoza, C/ Pedro Cerbuna 12, 50009 Zaragoza, Spain
}

\author{S. Bustingorry}
\affiliation{ 
Instituto de Nanociencia y Nanotecnolog\'ia (CNEA-CONICET), Nodo Bariloche, Av. Bustillo 9500 (R8402AGP), S. C. de Bariloche, R\'io Negro, Argentina
}
\affiliation{
Gerencia de F\'isica,  Centro Atómico Bariloche, Av. Bustillo 9500 (R8402AGP), S. C. de Bariloche, R\'io Negro, Argentina
}
\affiliation{ 
Aragon Nanoscience and Materials Institute (CSIC‐University of Zaragoza) and Condensed Matter Physics Department, University of Zaragoza, C/ Pedro Cerbuna 12, 50009 Zaragoza, Spain
}

\date{\today}

\begin{abstract}
Spin polarized currents originate a spin-transfer torque that enables the manipulation of magnetic textures.
Here we theoretically study the effect of a spin-polarized current on the magnetic texture corresponding to a chiral soliton lattice in a monoaxial helimagnet under a transverse magnetic field. 
At sufficiently small current density the chiral soliton lattice reaches a steady motion state with a velocity proportional to the intensity of the applied current, the mobility being independent of the density of solitons and the magnetic field.
This motion is accompanied with a small conical distortion of the chiral soliton lattice. At large current density the spin-transfer torque destabilizes the chiral soliton lattice, driving the system to a ferromagnetic state parallel to the magnetic field. We analyze how the deformation of the chiral soliton lattice depends on the applied current density. The destruction of the chiral soliton lattice under current  could serve as a possible erasure mechanisms for spintronic applications.
\end{abstract}

\maketitle

\section{\label{sec:intro}Introduction}

In magnetic systems where the antisymmetric Dzyaloshinskii-Moriya interaction (DMI) is present~\cite{Dzyal58,moriya1960new}, topological and chiral features emerge. The DMI interaction is the responsible of the stabilization of localized magnetic textures with chiral character, such as the skyrmion lattice~\cite{Muehlbauer09,Yu10,Yu11,Wilhelm11,kezsmarki2015neel,wu2020neel,Laliena18b} and single skyrmion state~\cite{Bogdanov94a,Bogdanov94b,Bogdanov99,sampaio2013nucleation}. In monoaxial helimagnets, such as CrNb$_3$S$_6$, CrTa$_3$S$_6$, CuB$_2$O$_4$, CuCsCl$_3$, Yb(Ni$_{1-x}$Cu$_x$)$_3$Al$_9$ and Ba$_2$CuGe$_2$O$_7$~\cite{Moriya82,Kousaka16,Roessli01,Adachi80,Ohara14,Matsumura17,Zheludev97,togawa2012chiral}, the DMI favors the rotation of the magnetization along a single chiral axis. In this case, analogously to the skyrmion lattice and single skyrmion in bulk or interfacial DMI systems, chiral soliton lattice (CSL)~\cite{Dzyal64, Miyadai83, izyumov1984modulated, togawa2012chiral, Kishine15, Togawa16, Laliena16a, Laliena16b, Laliena17a} and individual chiral solitons (CSs) can be stabilized~\cite{victor2020dynamics}. 

Both objects, the skyrmions and chiral solitons, present interesting magnetoresistive~\cite{hanneken2015electrical,Togawa13,Togawa15} and mobility~\cite{sampaio2013nucleation,iwasaki2013universal,victor2020dynamics} properties, with their particular imprint related to their structure and topological nature. These properties make them good candidates for spintronic devices~\cite{skyrmionicsroadmap2020}.
Besides the application to spintronic devices, new electromagnetic properties of magnetic textures are being explored based on the concept of emergent electrodynamics~\cite{schulz2012emergent,Nagaosa13}. 
It was theoretically predicted, and experimentally confirmed in the compound Gd$_3$Ru$_4$Al$_{12}$, that the spiral structure encountered in helimagnets can effectively work as an electromagnetic inductor~\cite{nagaosa2019emergent,yokouchi2020emergent}. This property of the spiral structure allows for the implementation of large inductances at small scales. 

The previously described potential technological applications motivate the study of the CS and CSL dynamics in monoaxial helimagnets under electric current. The response to external currents of the CSL has been theoretically studied in the linear response limit corresponding to small currents and weak fields~\cite{Kishine10,Tokushuku17}. The response of a single CS to external currents has been recently analyzed and it has been shown that the single soliton is destabilized and can be destroyed by large currents~\cite{victor2020dynamics}.  
Here, we study the response of the CSL in a wide range of currents and magnetic fields. 
We show that both the CSL and the single CS have the same mobility in the steady motion regime, 
and that the CSL is also destabilized with large currents. 
Our results are relevant within the field of chiral magnetism but also for the design of spintronic and electronic devices. 

The article is organized as follows: in Sec. \ref{sec:model} we introduce the model for a monoaxial chiral helimagnet under the effect of a spin-transfer torque, we present the main results on the CSL stability and subcritical dynamics in Sec. \ref{sec:sub}, we continue in Sec. \ref{sec:critic} with the study of the dynamical behavior in the supercritical regime, and in Sec. \ref{sec:phase_diagram} we study the $j-B$ phase diagram and the critical current at constant density of solitons. Finally we summarize our findings in Sec. \ref{sec:conclusions}.

%%%%%%%%%%%%%%%%%%%%%%%%%%%%%%%%%%%%%%%%%%

% model
\section{\label{sec:model}Micromagnetic model for a monoaxial helimagnet under external currents}

The time evolution of the magnetization field in a ferromagnet under current induced external torque is governed by the modified Landau-Lifshitz-Gilbert (LLG) equation:
\be
\label{ec:llg_corr}
\frac{\partial\bm{n}}{\partial t}=\gamma \bm{B}_{\mathrm{eff}}\times\bm{n}+\alpha \bm{n}\times\left(\frac{\partial\bm{n}}{\partial t}\right)  + \bm{\tau},
\ee
where $\alpha$ and $\gamma$ are the Gilbert damping and the gyromagnetic constant, respectively. The vector field $\bm{B}_{\mathrm{eff}}(\bm{r})=-\frac{1}{M_{\mathrm{S}}}\frac{\delta E}{\delta\bm{n}(\bm{r})}$ is the effective field derived from the energy functional $E$. The unimodular vector field $\bm{n}(\bm{r})=\bm{M}(\bm{r})/M_{\mathrm{S}}$ describes the local magnetization direction and $M_{\mathrm{S}}$ is the saturation magnetization.
The last term in Eq. (\ref{ec:llg_corr}), $\bm{\tau}$, is the spin-transfer torque due to the spin-polarized current and it is given by:
\be
\label{ec:stt_zhang_li}
\bm{\tau}=- (\bm{u}\cdot\nabla)\bm{n} + \beta \bm{n}\times(\bm{u}\cdot\nabla)\bm{n},
\ee
where $\bm{u} = - b_j \bm{j}$ and $b_j = \frac{P\mu_{\mathrm{B}}}{|e| M_{\text{S}}}$ with $P$ the polarization degree, $e$ the electron charge, and $\mu_{\mathrm{B}}$ the Bohr magneton. Notice that $\bm{u}$ points in the direction of the electron motion while the current density $\bm{j}$ points in the opposite direction. The first term is the reactive (adiabatic) torque and the second term is the dissipative (non-adiabatic) torque, whose strength is controlled by the nonadiabaticity coefficient $\beta$~\cite{Zhang04, Manchon19}. 

To describe a monoaxial chiral ferromagnet we consider a model that includes ferromagnetic exchange interactions, monoaxial DMIs and single-ion anisotropies, characterized by the stiffness constant $A$, the DMI strength constant $D$, and the anisotropy constant $K$, respectively.
Thus the magnetic energy functional is $E[\bm{n}]=\int d^{3}\bm{r} e(\bm{r})$, and the energy density $e(\bm{r})$ is given by
\be
\label{eq:ener_laliena}
e(\bm{r})=A\sum_{i}\left(\partial_{i}\bm{n}\right)^2-D\bm{\hat{z}}\cdot\left(\bm{n}\times\partial_{z}\bm{n}\right)-K n_{z}^{2}-M_{\mathrm{S}}\bm{B}\cdot\bm{n},
\ee
where the index $i$ runs over $x,y,z$, the chiral axis is along $\bm{\hat{z}}$ and $\bm{B}$ is the external magnetic field. The effects of the dipolar interaction are effectively taken into account in the uniaxial anisotropy term, which is correct for magnetization fields that depend only on the $z$ coordinate, as those considered in this work.
The corresponding effective field in Eq. (\ref{ec:llg_corr}) reads:
\be
\label{eq:b_eff}
\bm{B}_{\mathrm{eff}}=\frac{2}{M_{\mathrm{S}}}\left[ A\nabla^{2}\bm{n}-D\bm{\hat{z}}\times\partial_{z}\bm{n}+Kn_z\bm{\hat{z}}+\frac{M_{\mathrm{S}}}{2}\bm{B}\right].
\ee

\begin{figure}[t!]
\includegraphics[width=8cm]{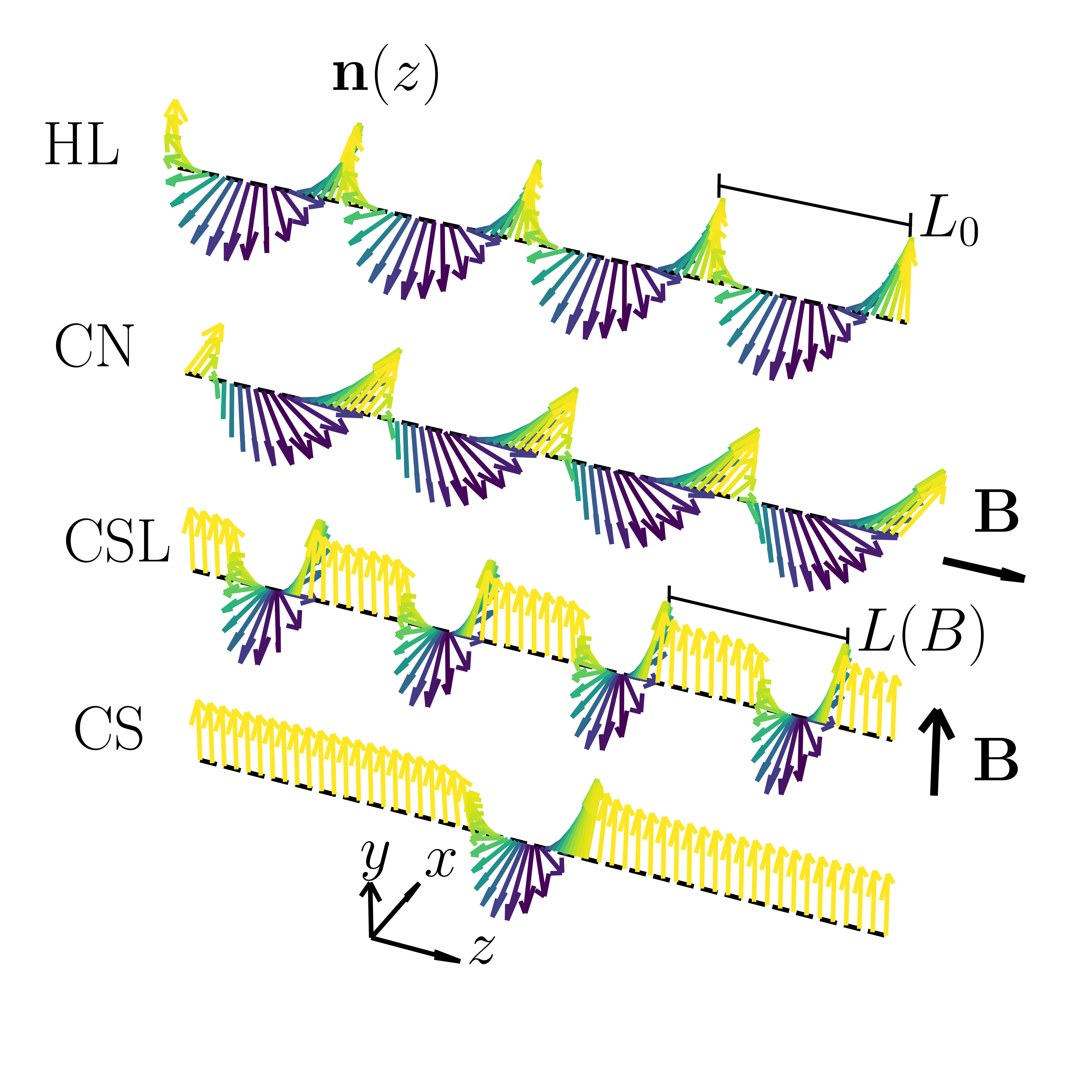}
\caption{The magnetization field for different configurations in a monoaxial chiral magnet: at zero magnetic field the configuration corresponds to the helical state (HL) with period $L_{0}$, for a magnetic field along the chiral axis the magnetization corresponds to the conical state (CN), if the magnetic field is applied in the direction perpendicular to the chiral axis the magnetic state corresponds to a chiral soliton lattice (CSL) which can be conceived as a regular arrangement of chiral solitons (CS). The color code represents the $n_{y}$ component: blue (yellow) for $n_{y} = -1$ (+1).
\label{fig:conf_sol}}
\end{figure}

The model just described possess a rich phenomenology. Without applied current and at zero magnetic field the magnetization forms a helical  structure (HL) with the propagation vector $\bm{q}_{0}$ aligned with the chiral axis (see Fig.~\ref{fig:conf_sol}). This means that the magnetization is contained within the $x-y$ plane but rotates around the $z$ axis.
If a magnetic field is applied along the chiral axis, the helical state features a conical deformation leading to a conical state (CN) as shown in Fig.~\ref{fig:conf_sol}. 
By increasing the magnetic field the system reaches a ferromagnetic state, with the magnetization pointing in the $z$ direction~\cite{Miyadai83, Ghimire13, Chapman14, Laliena16a, Laliena17a}.
Instead, if a magnetic field is applied in a direction perpendicular to the chiral axis, say $\bm{B}=B\bm{\hat{y}}$, the helical state is distorted and a CSL is formed (Fig.~\ref{fig:conf_sol}). The structure of the CSL can be transformed into that of the HL if the magnetic field is gradually reduced down to zero.
The density of solitons decreases with the external field $B$, so that the distance between consecutive solitons increases according to the relation \cite{togawa2012chiral,Dzyal64,izyumov1984modulated,kishine2005synthesis}
\be
\label{eq:l_teo}
\frac{L(B)}{L_0}=\frac{4\tilde{K}(k)\tilde{E}(k)}{\pi^{2}},
\ee
where $L_{0}=4\pi A/D$ is the period of the zero-field helical state, $\tilde{K}(k)$ and $\tilde{E}(k)$ are the complete elliptical integrals of the first and second kind, respectively, and $k$ solves the equation
\be
\label{eq:l_teo2}
\frac{k}{\tilde{E}(k)}=\sqrt{\frac{B}{B_{c}}}.
\ee

The model described by Eq. \eqref{eq:ener_laliena} applies to a wide range of monoaxial chiral helimagnets. In particular we shall consider $A=1.42 \un{pJ/m}$, $D=369 \un{\mu J/m^{2}}$, $K=-124 \un{kJ/m^3}$ and $M_{\mathrm{S}}=129 \un{kA/m}$, that reproduces the phenomenology of the CrNb$_{3}$S$_{6}$ compound~\cite{togawa2012chiral,Dzyal64,Miyadai83,izyumov1984modulated,kishine2005synthesis}. The zero-field helical pitch $L_{0}\approx48$ nm and the critical field $B_{c}\approx 230$ mT for the chiral soliton lattice-forced ferromagnet transition in a transverse magnetic field, are well described by the previous set of parameters~\cite{victor2020dynamics,Osorio2021}.

In the following, we shall study the effect of an external electric current  applied along the chiral axis when the system is subjected to a magnetic field  applied perpendicular to the chiral axis. Henceforth we thus consider a magnetic field along the $\bm{\hat{y}}$ direction, $\bm{B}=B\bm{\hat{y}}$.

\section{\label{sec:sub}Steady motion of the Chiral Soliton Lattice}

Since the norm of the magnetization $\bm{n}$ is constant there are only two degrees of freedom and it is useful to use the polar parametrization
\be
 \label{eq:polar_param}
 \bm{n} = -\sin \theta \sin \varphi \, \bm{\hat{x}} + \sin \theta \cos \varphi \, \bm{\hat{y}} + \cos \theta \, \bm{\hat{z}},
\ee
with the direction $\bm{\hat{z}}$ aligned with the chiral axis.

Steady solutions of the LLG equation, where a magnetic texture rigidly moves at a constant velocity, exist if there is an applied electric current which delivers a torque on the magnetization.
In this case the magnetic state is characterized by functions $\theta(w)$ and $\varphi(w)$ depending on $w = q_0(z-vt)$, with $v$ a constant velocity and $q_0 = D/2A$.
Setting the current to $\bm{j} = -j \bm{\hat{z}}$, the LLG equations in the steady state can be written in the form
\begin{widetext}
\bea
% \begin{gather}
  \theta^{\prime\prime} &=& (\varphi^{\prime\,2}-2\varphi^\prime + \kappa)\sin\theta\cos\theta - h_y\cos\theta\cos\varphi - \Omega\theta^\prime + \Gamma\sin\theta\varphi^\prime, \label{eq:stationary1} \\[4pt]
  \sin\theta\varphi^{\prime\prime} &=& h_y\sin\varphi - 2(\varphi^\prime-1)\cos\theta\theta^\prime - \Gamma\theta^\prime - \Omega\sin\theta\varphi^\prime, \label{eq:stationary2}
% \end{gather}
\eea
\end{widetext}
where $\kappa=K/Aq_0^2$ and $h_y=M_\mathrm{S}B/2Aq_0^2$. The primes indicate derivatives with respect to the $w$ variable. The parameters $\Omega$ and $\Gamma$ are given by
\be
\label{eq:Omega_Gamma}
 \Omega = \frac{\alpha}{v_0} \left( v - \frac{\beta}{\alpha} b_j j \right), \quad
 \Gamma = \frac{1}{v_0} \left(v - b_j j \right), 
\ee
with $v_0 = 2 \gamma A q_0/M_\mathrm{S}$.
When the current is applied to the CSL, the steady solution is expected to be also periodic and thus the steady equations are solved for $z$ within an interval of length equal to a period, $L$. This means $w\in[-w_L,w_L]$ with $w_L=q_0L/2$, and then $\varphi(w)$ and $\theta(w)$ satisfy the boundary conditions

\begin{gather}
 \varphi(-w_L) = 0, \;\; \varphi(w_L) = 2 \pi, \;\; \varphi'(w_L) = \varphi'(-w_L), \label{eq:BC1} \\
\theta(-w_L) = \theta(w_L), \;\;\; \theta'(w_L) = \theta'(-w_L). \label{eq:BC2}
\end{gather}
These conditions ensure, in a single period, a $2 \pi$ rotation of $\varphi$, periodicity of $\theta$ and continuity of their derivatives.

\subsection{Determination of the steady solutions \label{subsec:steady_bvp}}

Besides the model parameters and the applied magnetic field, Eqs.~\eqref{eq:stationary1} and \eqref{eq:stationary2} contain a priori two independent free parameters, $\Omega$ and $\Gamma$, or, 
equivalently, $j$ and $v$. The value of $j$ can be arbitrarily chosen since it corresponds to an external physical parameter which can be varied at will. 
However, we expect the velocity $v$, which has been introduced in the ansatz for the
steady state solution, to be determined by the applied current.
%However, we do not expect that the velocity $v$, which has been introduced in the ansatz for the steady state solution, can be freely chosen, independently of $j$. On the contrary, we expect that $v$ be determined by the applied current. 
This is indeed what happens, since the boundary value problem defined by Eqs.~\eqref{eq:stationary1} and \eqref{eq:stationary2} and the boundary conditions 
\eqref{eq:BC1} and \eqref{eq:BC2} has a solution \textit{only if} $\Omega=0$, as shown in appendix~\ref{sec:app-bvp}. In this way the current $j$ determines uniquely the steady state velocity $v$, which is given by

\be
\label{eq:v_vs_j} 
v=\frac{\beta b_{j}}{\alpha}j.
\ee

This means that the steady velocity has a linear dependence with the current density $j$, with a mobility $m = \beta b_j/\alpha$ which is independent of the density of solitons and of the applied field, but still depends on the Gilbert damping, the non-adiabaticity parameter, the saturation magnetization and the polarization degree of the current. 
Notice that the direction of velocity vector $\bm{v}$ is opposite to the direction of the current density $\bm{j}$.
Interestingly, the relation in Eq. \eqref{eq:v_vs_j} is the same as that found for the steady motion of a single CS in a monoaxial helimagnet \cite{victor2020dynamics} and of a domain wall in an anisotropic ferromagnet~\cite{Thiaville05}. Thus, it seems to be a universal feature of the one dimensional magnetic soliton dynamics.
Notice that if the condition in Eq. \eqref{eq:v_vs_j} holds, $\Gamma$ is proportional to the current density: $\Gamma=(\beta/\alpha-1)b_jj/v_0$.

For $\Omega=0$ the boundary value problem defined by Eqs. \eqref{eq:stationary1}, \eqref{eq:stationary2}, \eqref{eq:BC1}, and \eqref{eq:BC2} may have one or more solutions, or no solution (this happens if $j$ is large, see below). For given $j$ we characterize the solutions by the magnetization tilt angle at the boundary \footnote{The reason to choose the magnetization tilt angle at the boundary instead of, for instance, at the cell center, is related to the method of solution of the boundary value problem (see appendix A).}, $\theta_L = \theta(-w_L) = \theta(w_L)$,  which encodes conical deformations of the magnetic configuration.

\begin{figure}[t!]
\includegraphics[width=8cm]{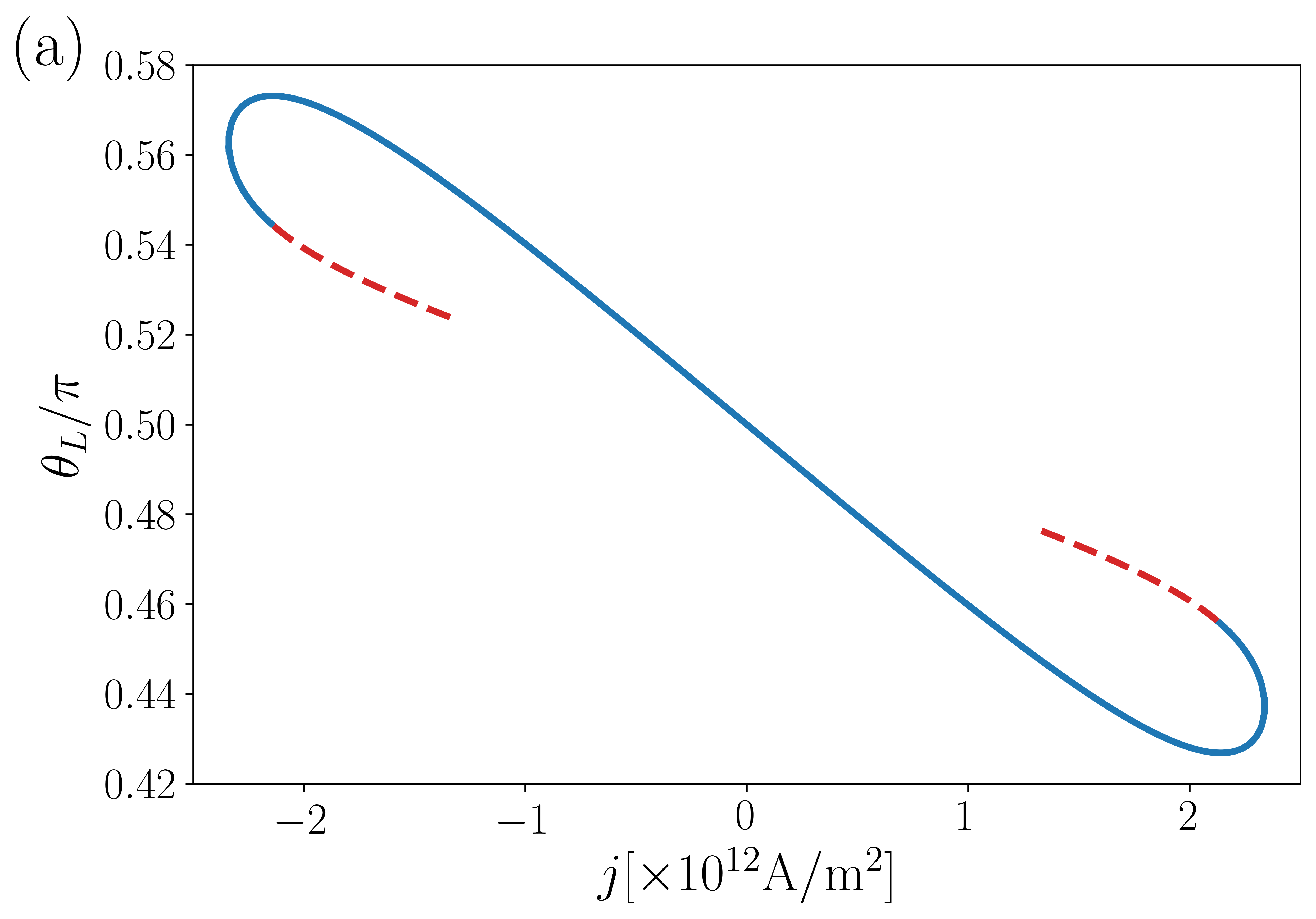}
\includegraphics[width=8cm]{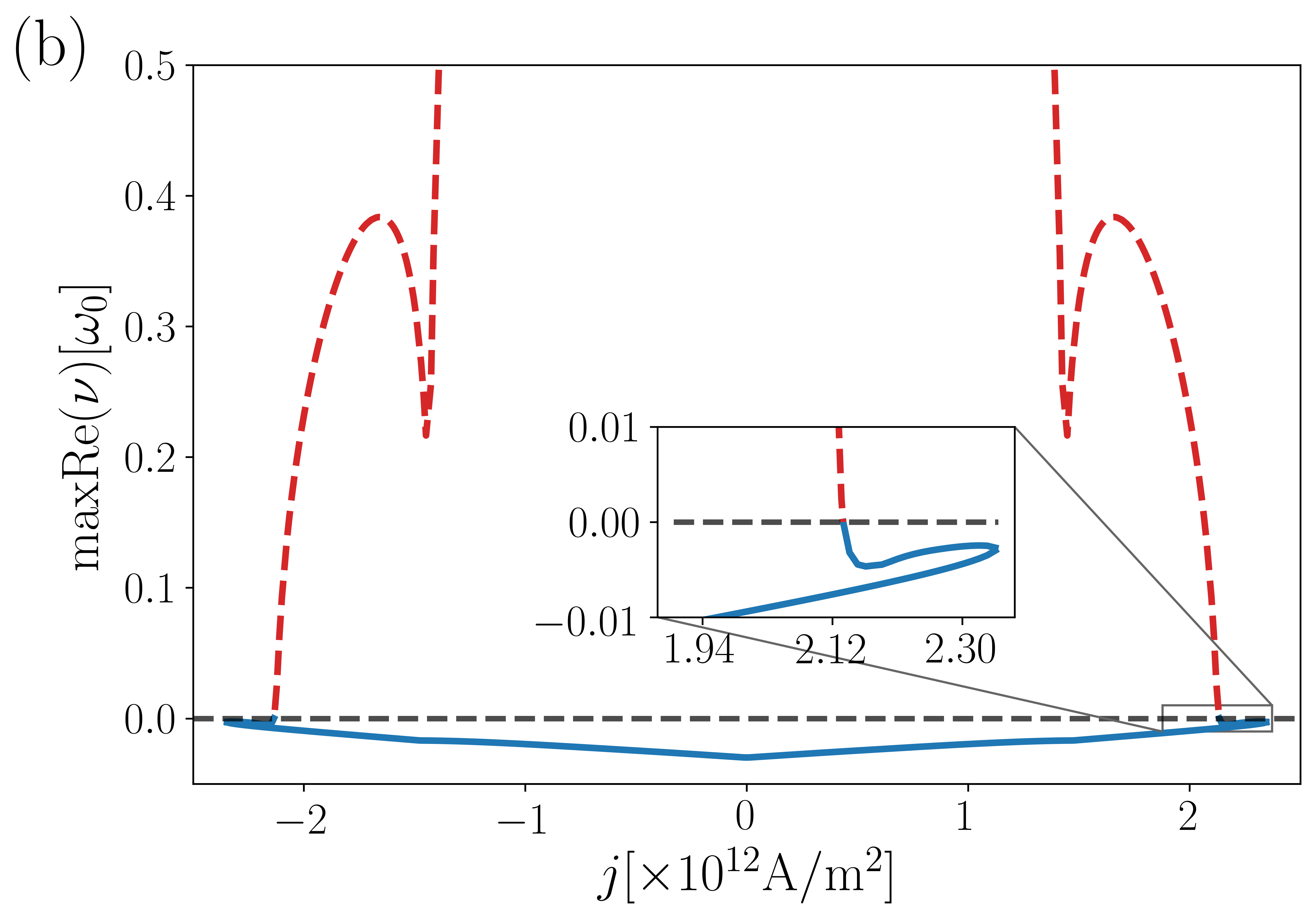}
\caption{(a) $\theta_L$ and (b) $\max\mathrm{Re}(\nu)$ (in units of $\omega_0$, see Appendix~\ref{sec:app-stability}) as a function of $j$ for $B = 50\un{mT}$. The stable branch of $\theta_L$ corresponds to $\max\mathrm{Re}(s) < 0$ and is indicated with a continuous blue line. Unstable branches are indicated with dashed red lines. For this value of the external field, there are no solutions beyond $j_c\approx2.34\times10^{12}\un{A/m^{2}}$. The inset in (b) shows that within the range $2.12\times10^{12}\un{A/m^{2}}\lesssim |j| \lesssim 2.34\times10^{12}\un{A/m^{2}}$ two stable solutions are found (corresponding to two different values of $\theta_{L}$ in (a)).
\label{fig:critic-sol}}
\end{figure}

For given $B$ and low values of $|j|$ there is only one solution, but at high enough $|j|$ a second solution appears. The two solutions merge at a critical value of $|j|$, denoted by $j_c$, beyond which the boundary value problem with $\Omega=0$ has no solution. 
As an example, Fig.~\ref{fig:critic-sol}(a) shows the values of $\theta_L$ as a function of $j$ for $B=50\un{mT}$, with a density of solitons corresponding to the equilibrium CSL at zero current, that is, with $L$ obtained from $B$ by Eq. \eqref{eq:l_teo}.
In this case, the value   $j_{c}\approx2.34\times10^{12}\un{A/m^{2}}$ is obtained.
The continuous blue line corresponds to stable solutions while the solutions indicated by broken red lines are unstable, as detailed in the following.

To analyze the stability of the steady solutions we study the dynamics of perturbations about them. Let $\bm{n}_0$ be a steady state and let a perturbation around this state be given by 
\be
\label{eq:pert}
\bm{n} = \bm{n}_0 + \xi_1 \bm{e}_1 + \xi_2 \bm{e}_2, 
\ee
where $\bm{e}_1$ and $\bm{e}_2$ are two orthonormal vectors perpendicular to $\bm{n}_0$, and $\xi_1$ and $\xi_2$ are the amplitudes of the perturbations. The perturbations $\xi_1$ and $\xi_2$ are functions of the three coordinates $x$, $y$, $z$, and of time, $t$, while the vectors $\bm{n}_0$, $\bm{e}_1$, and $\bm{e}_2$ are functions of the single variable $w=q_0(z-vt)$, where $v$ is given by Eq. \eqref{eq:v_vs_j}.
Inserting the form of the magnetization given by Eq. \eqref{eq:pert} into the LLG equation and linearizing it in $\xi_1$ and $\xi_2$ we obtain a linear equation for the dynamics of the perturbations.
Defining the two component column vector $\xi = (\xi_1, \xi_2)^T$, where the superscript $T$ stands for matrix transpose, the linearized LLG equation relates the time derivative of $\xi$ to a linear second order differential operator acting on $\xi$. The linear operator involves only spatial derivatives and its coefficients are functions only of $w$. Hence, it is convenient to perform a change of variables and consider $\xi$ a function of $t$, $x$, $y$ and $w$. In this form we obtain the equation
\be
\partial_t \xi = \mathcal{S} \xi, \label{eq:lin_LLG}
\ee
where the coefficients of the linear differential operator $\mathcal{S}$, which is given in Appendix~\ref{sec:app-stability}, depend only on $w$. With the ansatz $\xi = \eta e^{\nu t}$, where $\eta$ is a function of $x$, $y$, and $w$, the evolution equation is reduced to the eigenvalue problem $\mathcal{S}\eta = \nu\eta$. The steady state is stable if and only if all eigenvalues $\nu$ of $\mathcal{S}$ have non positive real part.

Figure~\ref{fig:critic-sol}(b) shows the maximum of the real part of the eigenvalues of $\mathcal{S}$ corresponding to the steady solutions of Fig.~\ref{fig:critic-sol}(a). 
Some details on the computations are given in the Appendix~\ref{sec:app-stability}.
We see that the blue branch of Fig.~\ref{fig:critic-sol}(a) represents the values of $\theta_L$ that correspond to stable steady solutions, while the steady solutions corresponding to the dashed branches are unstable. 
In the range $2.12\times10^{12}\un{A/m^{2}}\lesssim |j| \lesssim 2.34\times10^{12}\un{A/m^{2}}$, we find two possible stable solutions, as $\theta_{L}$ is not single valued and the corresponding eigenvalues have negative real part (see inset in Fig.~\ref{fig:critic-sol}(b)).
In this case, which of the two possible stable solutions is reached will depend on the initial condition. In our numerical simulations we use the CSL as the initial state and we always observe the solution corresponding to the maximum deviation from the $x-y$ plane, i.e. with $\max(|\theta_L - \pi/2|)$, corresponding to the lower(upper) blue section for positive(negative) $j$ values in Fig.~\ref{fig:critic-sol}(a).

In conclusion, steady motion states exist only if the applied current density is lower than a critical current $j_c$, which depends strongly on the applied magnetic field and on the density of solitons (see Sec. \ref{sec:phase_diagram}).

\subsection{Steady velocity-current response}

The stable steady solutions are reproduced by micromagnetic numerical simulations: a steady motion state is obtained after a short transient if a polarized electric current along the chiral axis is applied to a system which is initially at equilibrium, provided the applied current density is lower than a certain critical value.

\begin{figure}[t!]
\includegraphics[width=8cm]{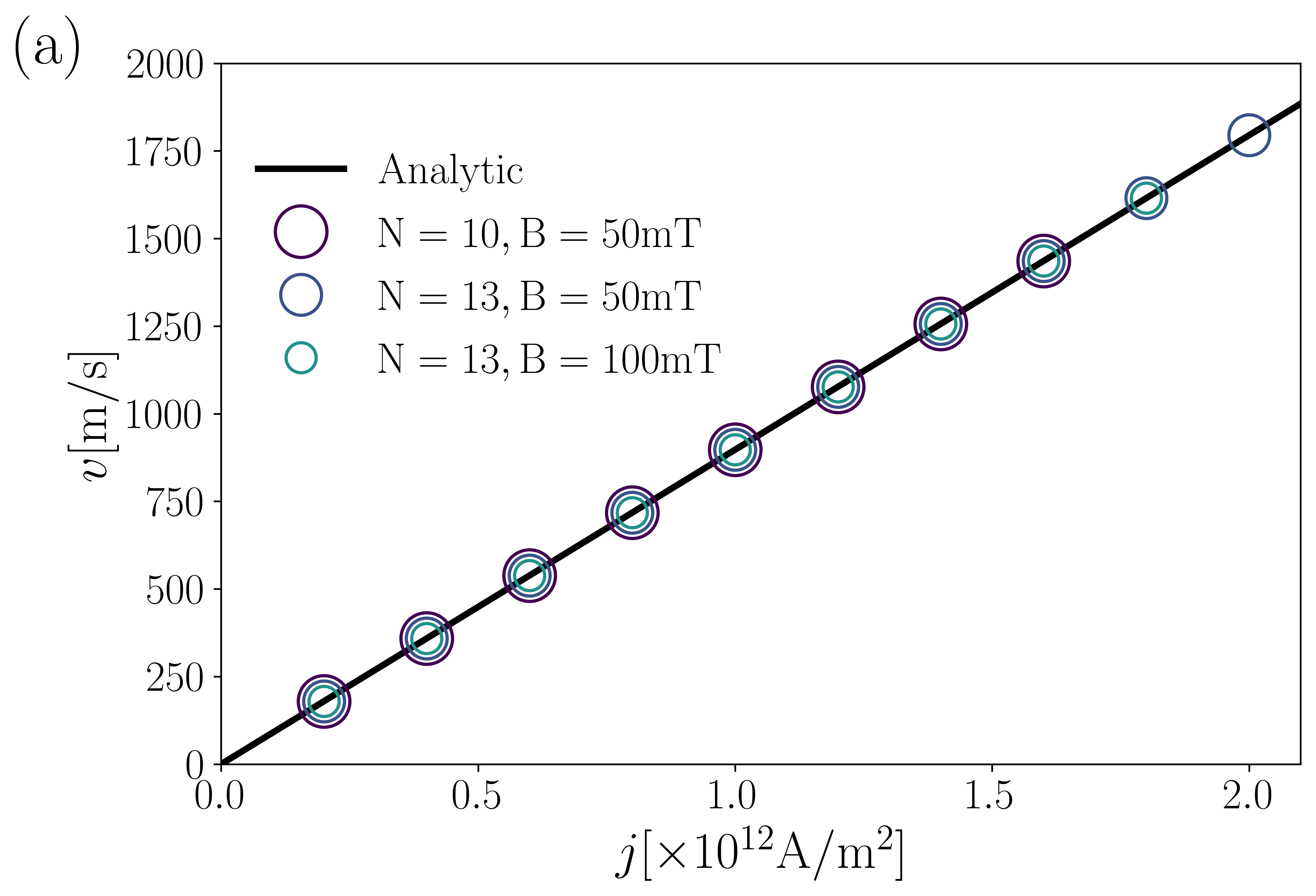}

\includegraphics[width=8cm]{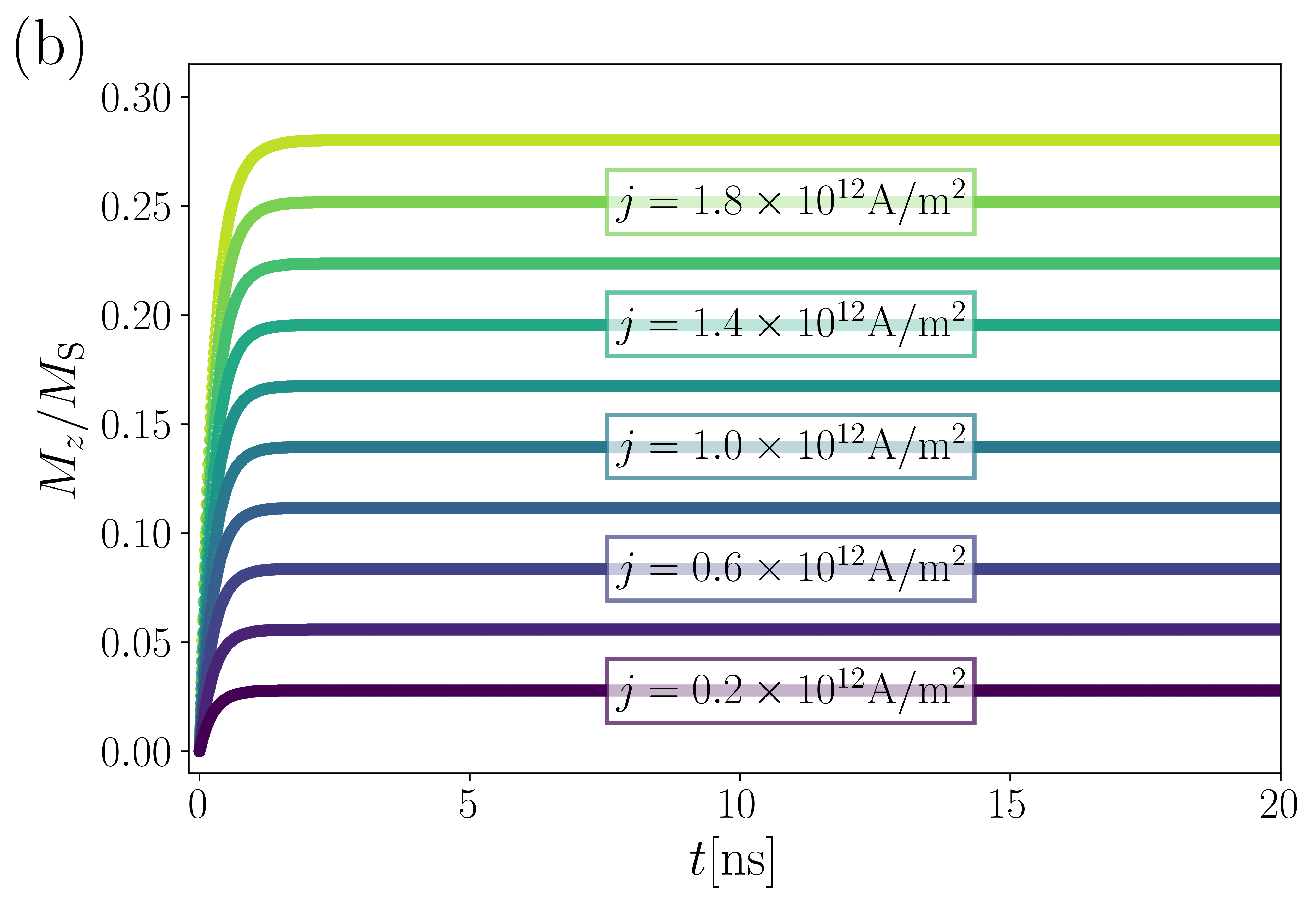}
\caption{(a) The CSL velocity for different number of CSs and at different values of magnetic field: $N=10$ and $B=50\un{mT}$, $N=13$ and $B=50\un{mT}$ and $N=13$ and $B=100\un{mT}$. The black line represents the analytical result for the velocity given by Eq.~\eqref{eq:v_vs_j}. % of a single CS.
(b) The magnetization along the chiral axis as a function of time for different square pulses of current of intensities $j$ and $B=50\un{mT}$.
\label{fig:v_vs_j}}
\end{figure}

We use the MuMax3 code and implement a monoaxial DMI interaction~\cite{MuMax3,Leliaert18,victor2020dynamics}. Parameter values for CrNb$_3$S$_6$ (as mentioned in Sec.~\ref{sec:model}) were used in a one-dimensional system of size $R =500\un{nm}$, with a mesh comprised of 500 cells of length $\Delta R=1\un{nm}$, and we set $\alpha=0.01$ and $\beta=0.02$ for the Gilbert damping in Eq. \eqref{ec:llg_corr} and the non-adiabaticity constant in Eq. \eqref{ec:stt_zhang_li}, respectively. We perform our simulations using periodic boundary conditions and keeping the number of chiral solitons constant at a given value $N$.
The velocity of the CSL can be obtained from the simulations using the autocorrelation $\langle \bm{n}(z,0)\cdot\bm{n}(z,t)\rangle$ where $\langle\cdots\rangle=\frac{1}{R}\int_{0}^{R}\cdots dz$. From the Fourier transform of the time-dependent autocorrelation function, and using the lowest non-zero frequency $\nu_{1}$, we get the CSL velocity as $v=\frac{\nu_{1}R}{2\pi N}$ (see Appendix ~\ref{sec:app-autocorrelation}). The results of the velocity as a function of the current are shown in Fig.~\ref{fig:v_vs_j}(a), indicating an extremely good agreement between the stationary solution and numerical simulations of the full LLG equations. The fact that the velocity does not depend on the solitons' density, controlled by the external magnetic field, gives room to work in a wide field range without modifying the dynamical properties of the CSL.

\subsection{Current induced CSL deformation}
\label{subsec:def}

\begin{figure}[t!]
\includegraphics[width=8cm]{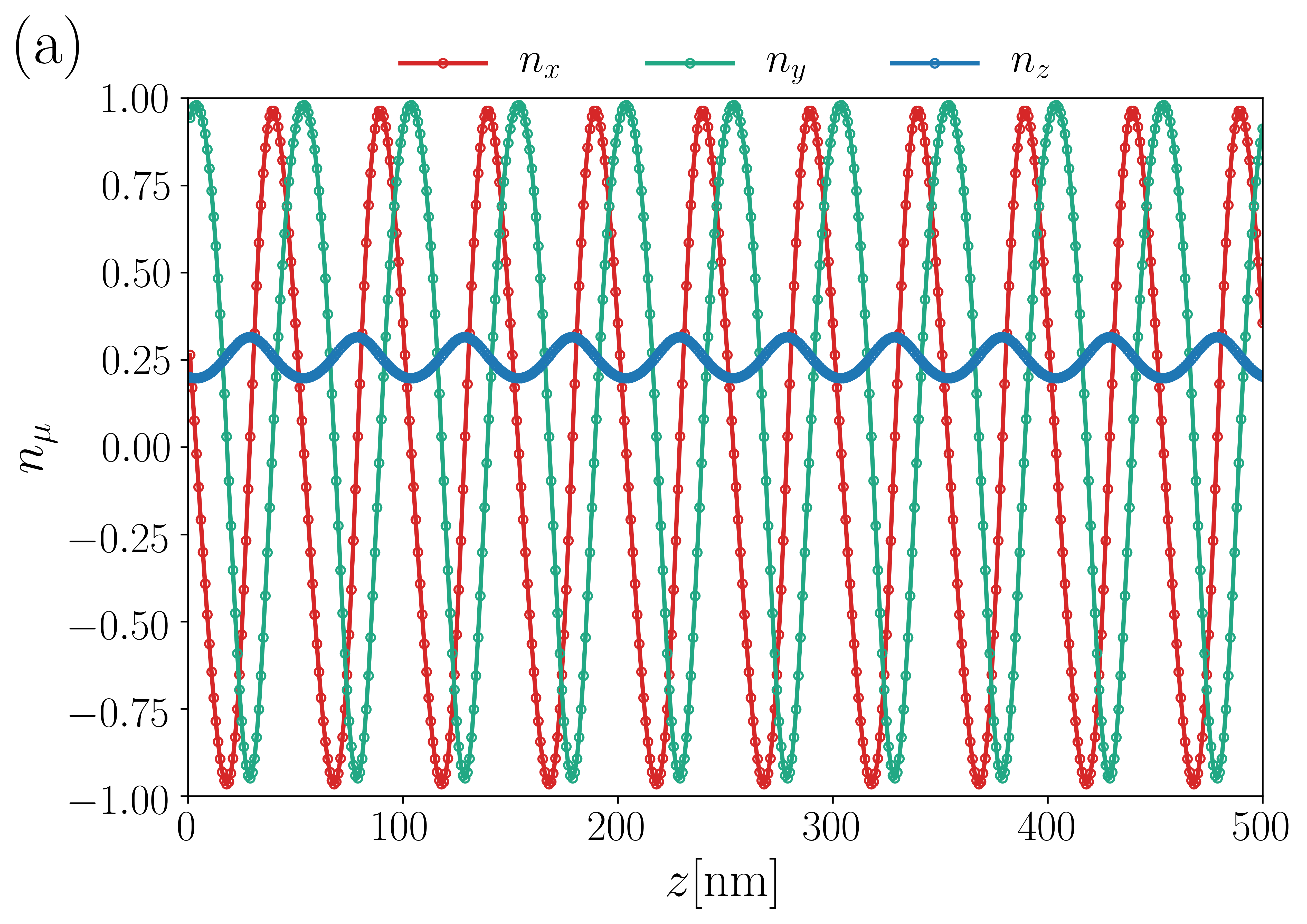}
\includegraphics[width=4cm]{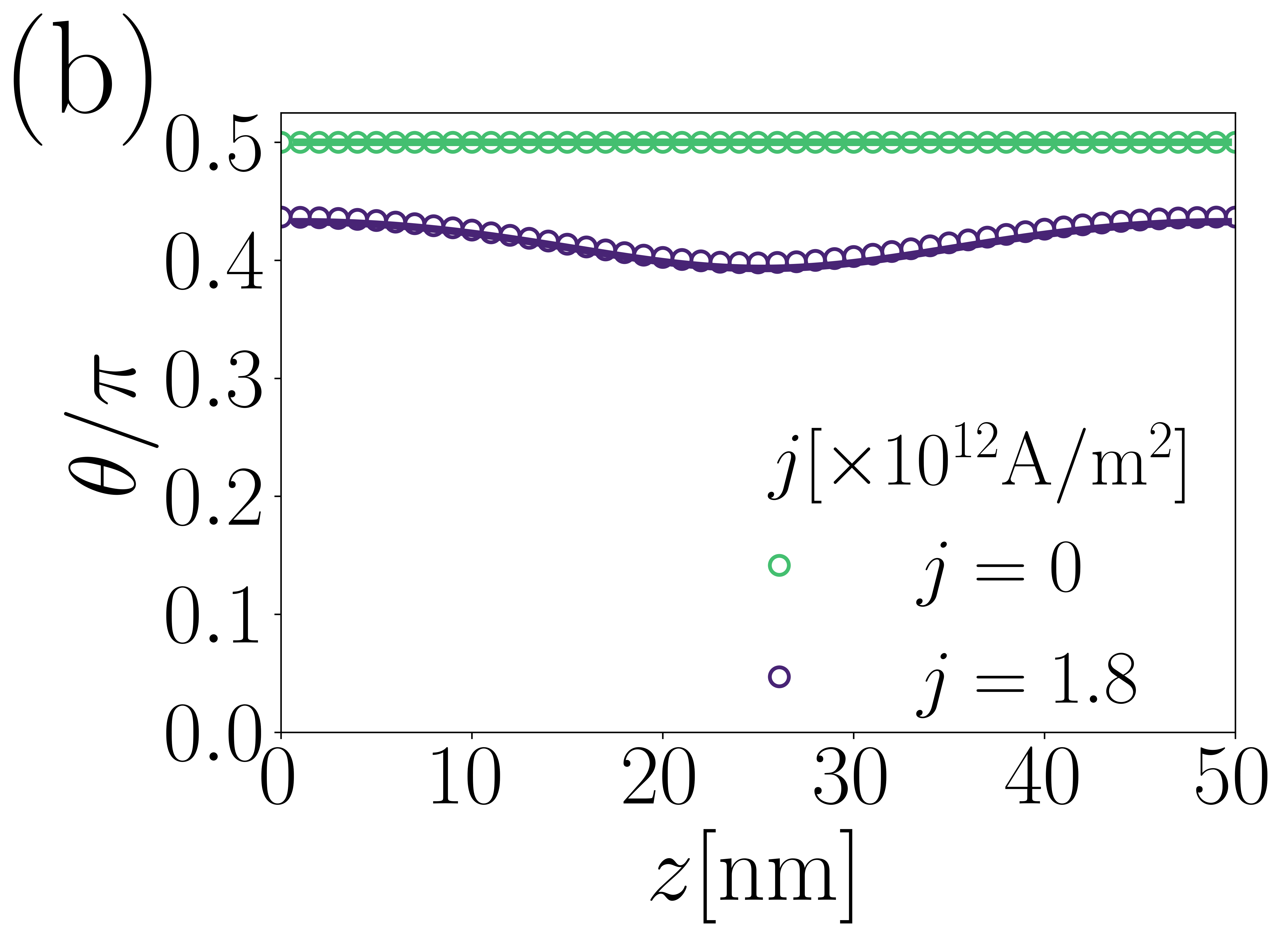}
\includegraphics[width=4cm]{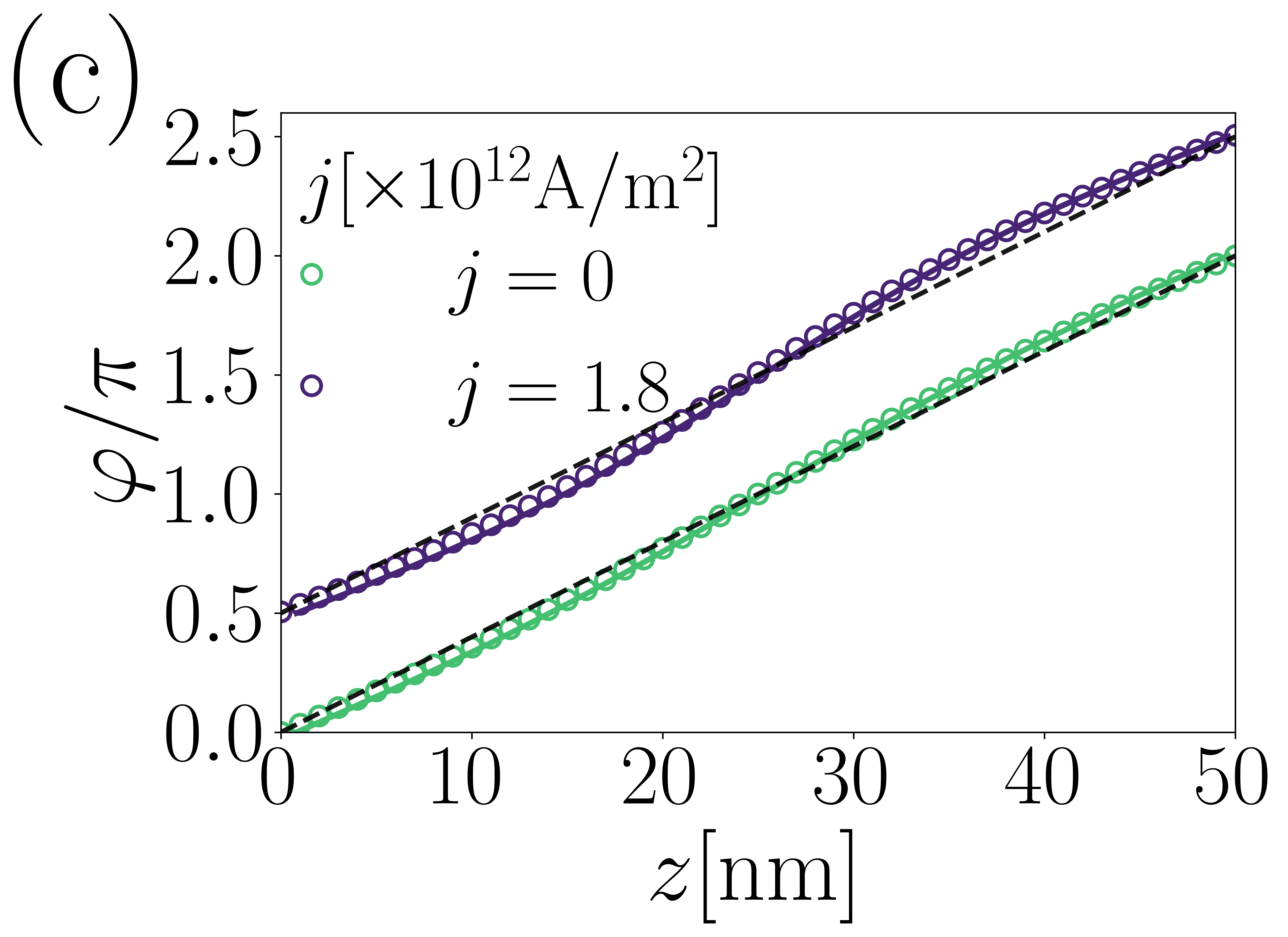}
\includegraphics[width=3.3cm]{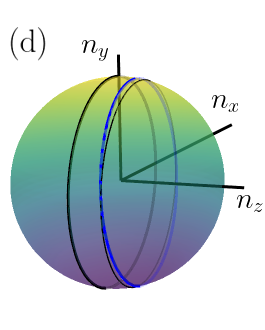}
\includegraphics[width=3.8cm]{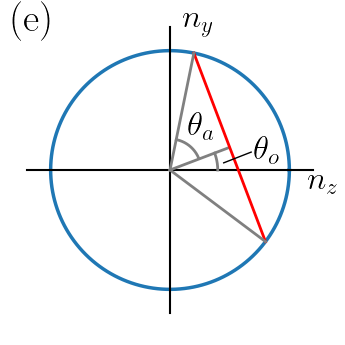}
\caption{
(a) A snapshot of the magnetization field along the sample after the steady motion is reached ($B=50$ mT and $j=1.8\times10^{12}$ A/m$^2$).
(b) Polar angle $\theta(z)$ within one period of the CSL, with a pitch $L=50\un{nm}$. 
(c) Rotation angle $\varphi(z)$ indicating one complete turn in a CSL period. The curve corresponding to $j=1.8\times10^{12}$ A/m$^2$ was displaced in order to present the results more clearly. The dotted lines serve as a guide for the eye and emphasize the difference between the cases with and without applied current. In (b) and (c) the circles represent the results from the micromagnetic simulations while the solid lines are the solutions for the boundary value problem in Eqs. \eqref{eq:stationary1}, \eqref{eq:stationary2}, \eqref{eq:BC1} and \eqref{eq:BC2}.
%\eqref{eq:stationary1}-\eqref{eq:BC1}.
(d) A spherical plot representing the magnetization field over the Bloch sphere. The thick black line represents the CSL before the current is applied. The thin black line represents the conical state for $B=0\un{mT}$ when the current is applied. The blue line represents the magnetization field in (a). The sphere represents the Bloch sphere spanned by the set of vectors $|\bm{n}|=1$ and the color code (blue-yellow) represents the value of $n_y$: blue (yellow) corresponds to $n_{y}=-1$ $(+1)$.
(e) Projection of the conical distortion in the $y-z$ plane. The orientation and opening angles, $\theta_o$ and $\theta_a$, characterizing the cone are indicated.
\label{fig:mz_nxnynz_esf}} 
\end{figure}

As shown in Fig.~\ref{fig:v_vs_j}(b), where the $z$ component of the net magnetization is presented, numerical simulations show that the stationary solutions are reached after a transient time of the order of a few nanoseconds. This results correspond to a case with $B = 50 \un{mT}$ and different intensities of the current $j$. It is also important to mention that besides the translation motion of the magnetic texture, the effect of the current involves a deformation of the original CSL into a state with cone-like profile, leading to a net magnetization along the chiral axis, as shown in Fig.~\ref{fig:v_vs_j}(b). At zero magnetic field, the current drives the system to a conical state analogous to the state observed in a cubic helimagnet under the same conditions~\cite{Goto08,masell2020manipulating,masell2020combing}. In this case the distortion is characterized by a uniform component of the magnetization field along the propagation vector $\bm{q}_0$. However, when a transverse magnetic field is applied, the magnetization component parallel to $\bm{q}$ is not uniform but exhibits a modulation along the system.
Figure~\ref{fig:mz_nxnynz_esf}(a) shows how the magnetization components are periodically varying along the $z$ coordinate, as found using micromagnetic simulations for $B=50\un{mT}$ and applying a current $j=1.8\times10^{12}\un{A/m^{2}}$.
The distortion of the CSL is described by the form of $\theta(w)$ and $\varphi(w)$ within one period. Figures~\ref{fig:mz_nxnynz_esf}(b) and \ref{fig:mz_nxnynz_esf}(c) compare the steady solutions obtained by solving the boundary value problem and by the micromagnetic simulations for $B=50\un{mT}$. 
A good agreement between both results is observed.

Let us discuss the form of the CSL distortion in the steady motion state.
In absence of current, $j=0$, the polar angle {has a constant value} $\theta(w) = \pi/2$, which means that the magnetization lays in the $x-y$ plane. 
If a current is applied,
$\theta(w)$ oscillates between a maximum value for $z = 0 ,L$ (i.e. $w = \pm w_{L}$) and a minimum value at $z=L/2$ (i.e. $w=0$), as can be appreciated in Fig.~\ref{fig:mz_nxnynz_esf}(b).
This means that the tilting of the magnetization towards the chiral axis is maximum at the center of the soliton, i.e. when $n_y$ is minimum, and it is minimum when $n_y$ takes its maximum value.
The variation of the angle $\varphi(w)$ indicates how the magnetization field performs the $2 \pi$ rotation, and depends on the applied current and field as shown in Fig.~\ref{fig:mz_nxnynz_esf} (c). 

The distortion of the steady moving CSL can be recast as a conical deformation, akin the one observed when a magnetic field in the $z$ direction is considered~\cite{Laliena17a,Yonemura17}. The opening of the cone depends on the intensity of the current. Large values of $j$ tend to shrink the cone, and, as a consequence, the value of the net magnetization along the chiral axis grows approximately linearly with the intensity of the current as shown in Fig. \ref{fig:v_vs_j}(b). In this case $\theta(w) < \pi/2$, indicating a conical deformation pointing in the $z$ direction.
It is instructive to represent the magnetization field over the Bloch sphere as in Fig. \ref{fig:mz_nxnynz_esf}(d). From this figure it is possible to recognize the effect of the current on the structure of the CSL: its profile changes from a planar (thick black line) to a conical section (thin black and thick blue lines) when a current density is applied. For $B=0\un{mT}$ the cone axis is aligned with the $z$ direction (thin black) whilst for non zero $B$ the orientation of the axis of the
conical distortion slightly departs from the $z$ direction (thick blue).

\begin{figure}[t!]
\includegraphics[width=8cm]{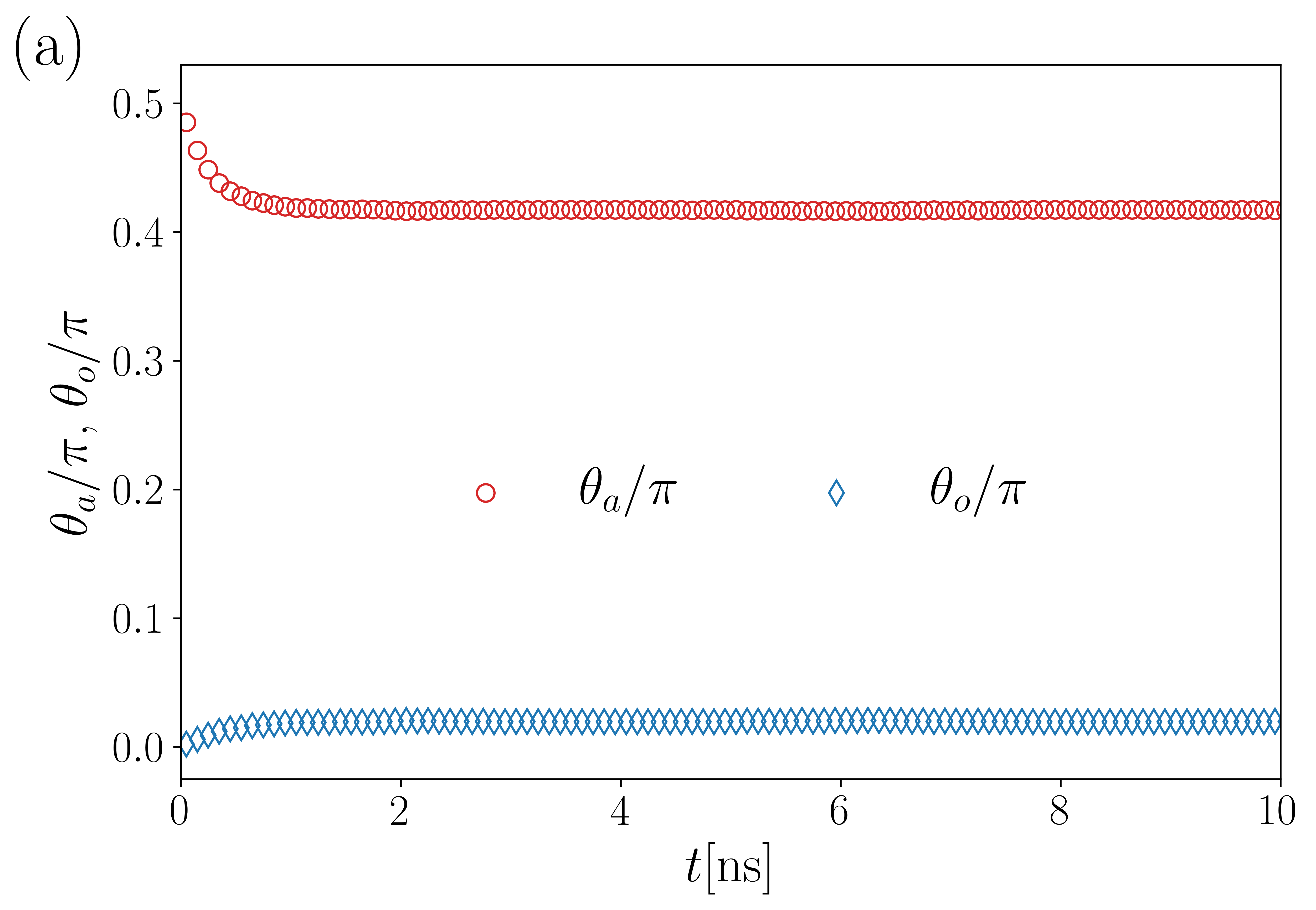}
\includegraphics[width=8cm]{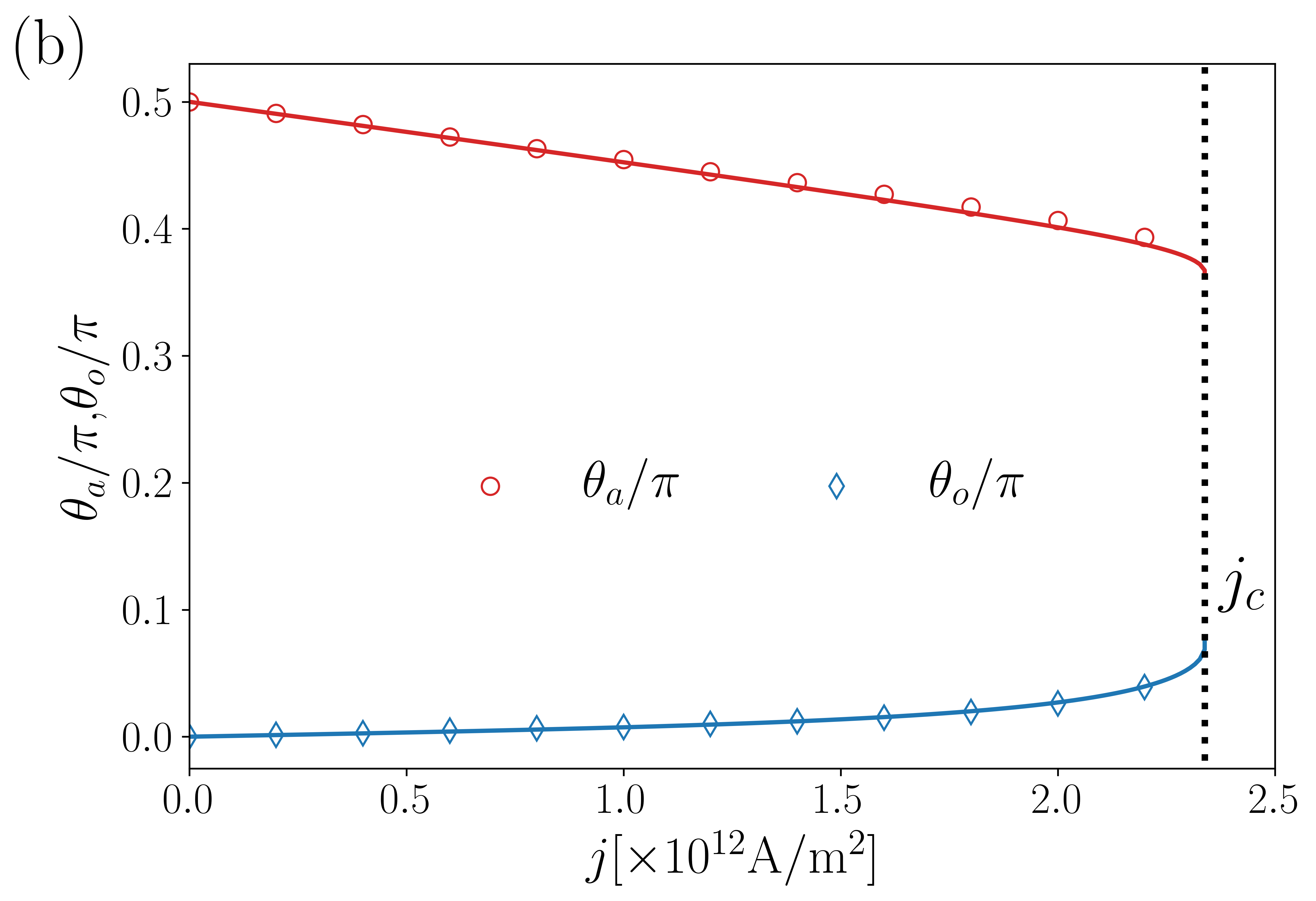}%
\caption{ Characteristics of the conical distortion for small currents.
(a) Time evolution of the orientation and opening angles, $\theta_o$ (blue diamonds) and $\theta_a$ (red circles), in the subcritical regime for $j=1.8\times10^{12}\un{A/m^2}$ and $B=50\un{mT}$. 
(b) The angles $\theta_{o}$ (in blue) and $\theta_{a}$ (in red), in the steady state, as a function of the current intensity for $B=50\un{mT}$. The circles and diamonds are the results from the micromagnetic simulations and the solid lines are the results obtained from the solution of the boundary value problem. The dotted black line signals the critical current $j_{c}\approx2.34\times10^{12}\un{A/m^{2}}$ for $B=50\un{mT}$.
\label{fig:angs}}
\end{figure}

Since the current deforms the CSL and turns its profile into an oriented-cone, key features of the magnetization dynamics can be characterized by two angles that we call $\theta_{o}$, providing information about the orientation of the cone, and $\theta_{a}$, representing the opening angle of the cone (see Fig. \ref{fig:mz_nxnynz_esf}(e)). Whenever $\theta_o >0$ the $2 \pi$ rotation of the magnetization is around the direction defined by $\theta_o$, and the cone is not perfectly oriented with the chiral axis.
Figure~\ref{fig:angs}(a) presents micromagnetic simulation results showing that $\theta_a$ (red circles) and $\theta_o$ (blue diamonds) reach a steady value. 
It can be observed that $\theta_{o}$ grows from zero (the axis of the cone coincides with the chiral axis) to a finite value in the steady regime, that is, the axis of the cone departs from the chiral axis. On the other hand, the opening angle $\theta_{a}$ decreases with time, from $\pi/2$ to a finite value reached at the steady state. The values of $\theta_{a}$ and $\theta_{o}$ in the steady state as a function of the applied current are shown in Fig.~\ref{fig:angs}(b). We see that $\theta_{a}$ decreases while $\theta_{o}$ increases with $j$. It is important to note that $\theta_{a}$ takes a finite value when $j$ reaches $j_c$, i.e. the critical regime is reached before the cone closes. The numerical results (symbols) and the analytical results (solid lines) are in perfect agreement. A similar phenomenology appears in the helical state of cubic noncentrosymmetric ferromagnets \cite{masell2020manipulating}.

\section{\label{sec:critic}Destruction of the chiral soliton lattice and transient dynamics beyond the critical current}

The steady states described in Section~\ref{sec:sub} are only reached if $j$ is below a critical current, since steady solutions of the LLG equation exist only if $j < j_c$ as indicated in Fig.~\ref{fig:critic-sol}. When $j > j_c$ the CSL is destabilized and the system is driven to a different state. 

\begin{figure}[t!]
\includegraphics[width=8cm]{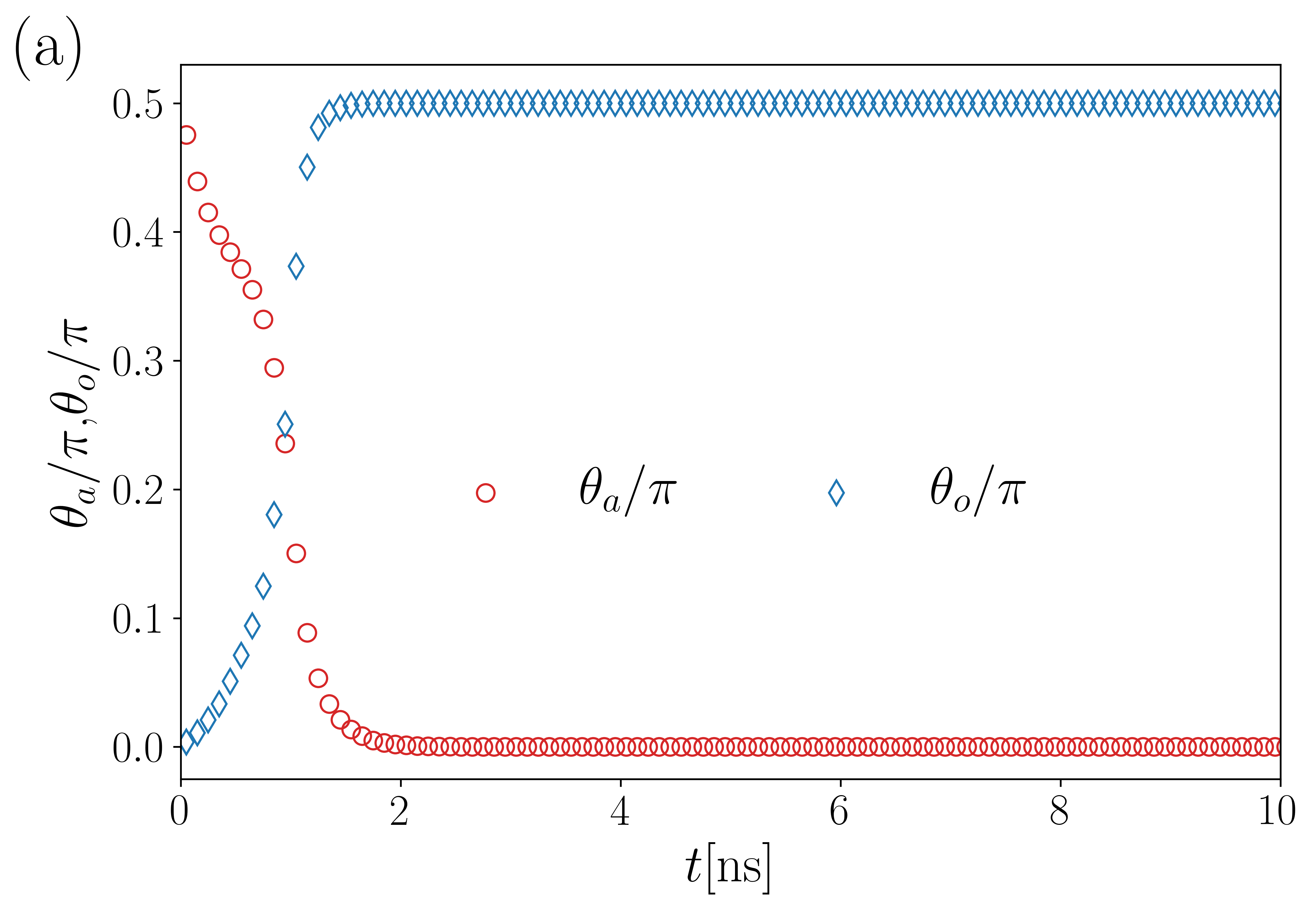}
\includegraphics[width=6.3cm]{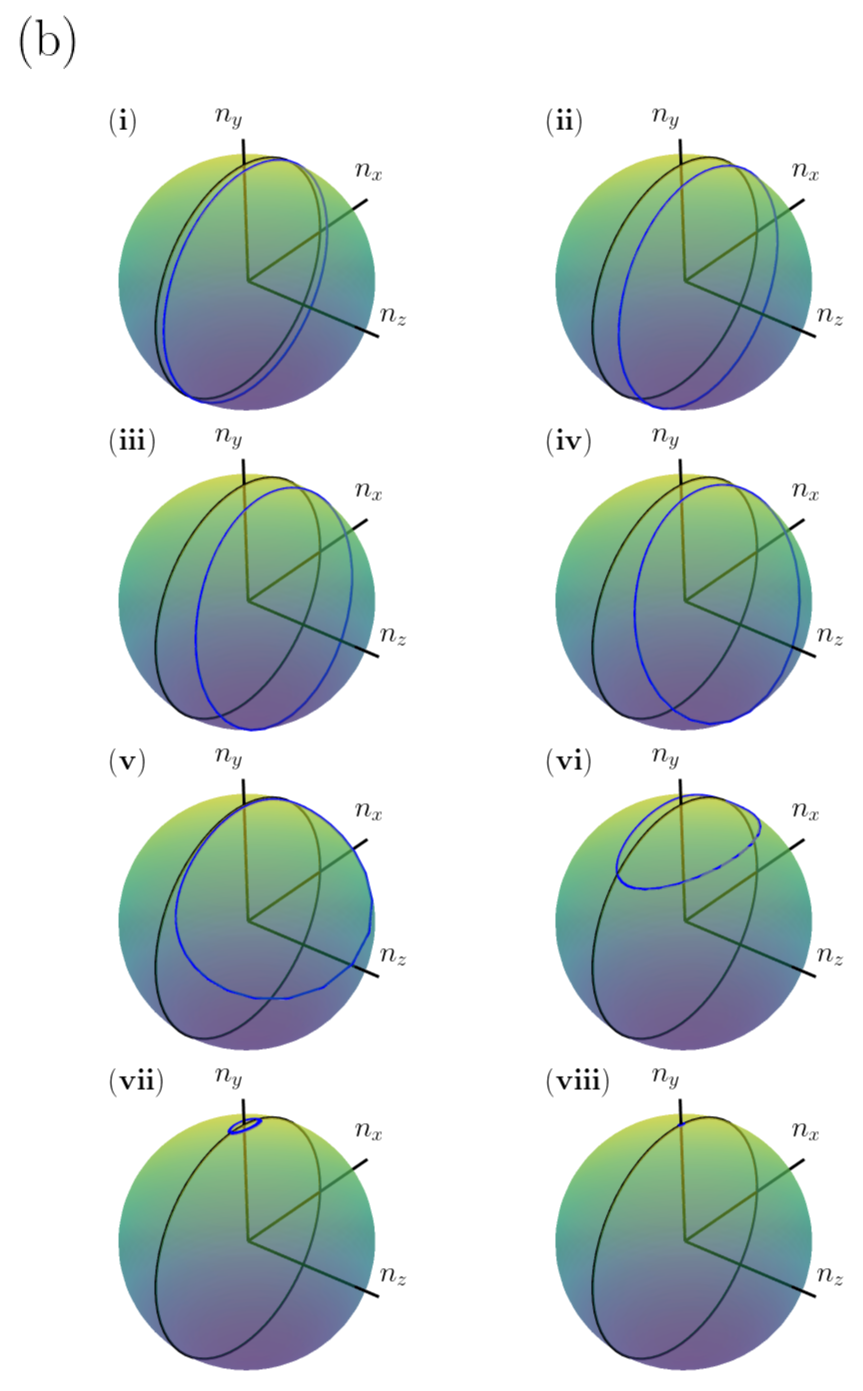}
\caption{
Destruction of the CSL in the supercritical current regime.
(a) Time evolution of the orientation and opening angles, $\theta_o$ (blue diamonds) and $\theta_a$ (red circles), in the supercritical regime for  $j = 3 \times 10^{12} \un{A/m^2}$. 
(b) Representation of the magnetization field (on the Bloch sphere) at selected times after the application of the density current pulse corresponding to (a): 
i) $t=0.05\un{ns}$,
ii) $t=0.20\un{ns}$,
iii) $t=0.45\un{ns}$,
iv) $t=0.70\un{ns}$,
v) $t=0.95\un{ns}$,
vi) $t=1.10\un{ns}$,
vii) $t=1.45\un{ns}$,
viii) $t=1.80\un{ns}$.
The black circle represents the initial state at $t=0\un{ns}$ and the blue line represents the magnetization at each time.
\label{fig:esfstar}}
\end{figure}

Although it is not expected to become an accurate description for large distortions, it is still insightful to describe the magnetization texture as an oriented cone. The time evolution of the orientation and opening angles obtained using micromagnetic simulations for $B = 50\un{mT}$ and for a current $j=3\times10^{12}\un{A/m^{2}}$, which is above $j_c$ ($j_{c}\approx2.34\times10^{12}\un{A/m^{2}}$ at $B=50\un{mT}$) are presented in Fig.~\ref{fig:esfstar}(a). The orientation angle $\theta_o$ (blue diamonds) starts increasing from zero and reaches the constant value $\theta_o = \pi/2$. Concomitantly, the value of the opening angle $\theta_a$ (red circles) decreases from $\pi/2$ to reach the constant value $\theta_a=0$. This means that the conical deformation initially oriented along the chiral axis rotates to the $y$ direction, whilst shrinking at the same time, and the final result is a ferromagnetic state ($\theta_a = 0$) oriented in the direction of the external magnetic field ($\theta_o = \pi/2$).

In Fig.~\ref{fig:esfstar}(b) we show a representation of the dynamical evolution of the magnetization field in the Bloch sphere for the current density and magnetic field values corresponding to Fig.~\ref{fig:esfstar}(a). It can be observed that after the application of the current the profile of the magnetization field can be pictured as a deformed cone with its axis pointing, approximately, along the chiral axis. The shape and orientation of this cone evolves with time and, after a while, the axis of the cone moves within the $y-z$ plane and its direction gradually departs from the chiral axis ($z$ axis) to finally lay along the direction of the magnetic field ($y$ axis), see Fig.~\ref{fig:esfstar}(b)i-vi. After this, the cross section of the cone starts shrinking to finally reach the ferromagnetic state along the magnetic field, see Fig.~\ref{fig:esfstar}(b)vii-viii. 
Notice that, as can be appreciated in Fig.~\ref{fig:angs}, the conical deformation does not fully close as $j$ approaches the critical current $j_c$ from below.
Moreover, notice also that once $\theta_o > \theta_a$ the magnetization texture winds around $\theta_o$, but the chiral axis is no longer contained within the cone defined by $\theta_o$ and $\theta_a$  [Figs.~\ref{fig:esfstar}(c)v-vi]. 
It is important to mention that after the destruction of the CSL the magnetic state can be described as a ferromagnetic state with small spatial fluctuations. As shown in Fig.~\ref{fig:esfstar}, the transition from the CSL to the ferromagnetic state occurs within a few nanoseconds. When the current is not too large ($j_{c}<j\leq j_{c}^{FM}$ with $j_{c}^{FM}$ the critical current for the ferromagnetic instability, discussed in Sec.~\ref{sec:phase_diagram}) the amplitude of these fluctuations decreases with time and the perfect ferromagnetic state is eventually reached.

To summarize the main results of this section we mention that for $j>j_{c}(B)$, but $j$ not too high, and a long enough pulse of current, the system reaches a ferromagnetic steady state, and the CSL exhibits a finite life time.

\begin{figure}[t!]
\includegraphics[width=8.0cm]{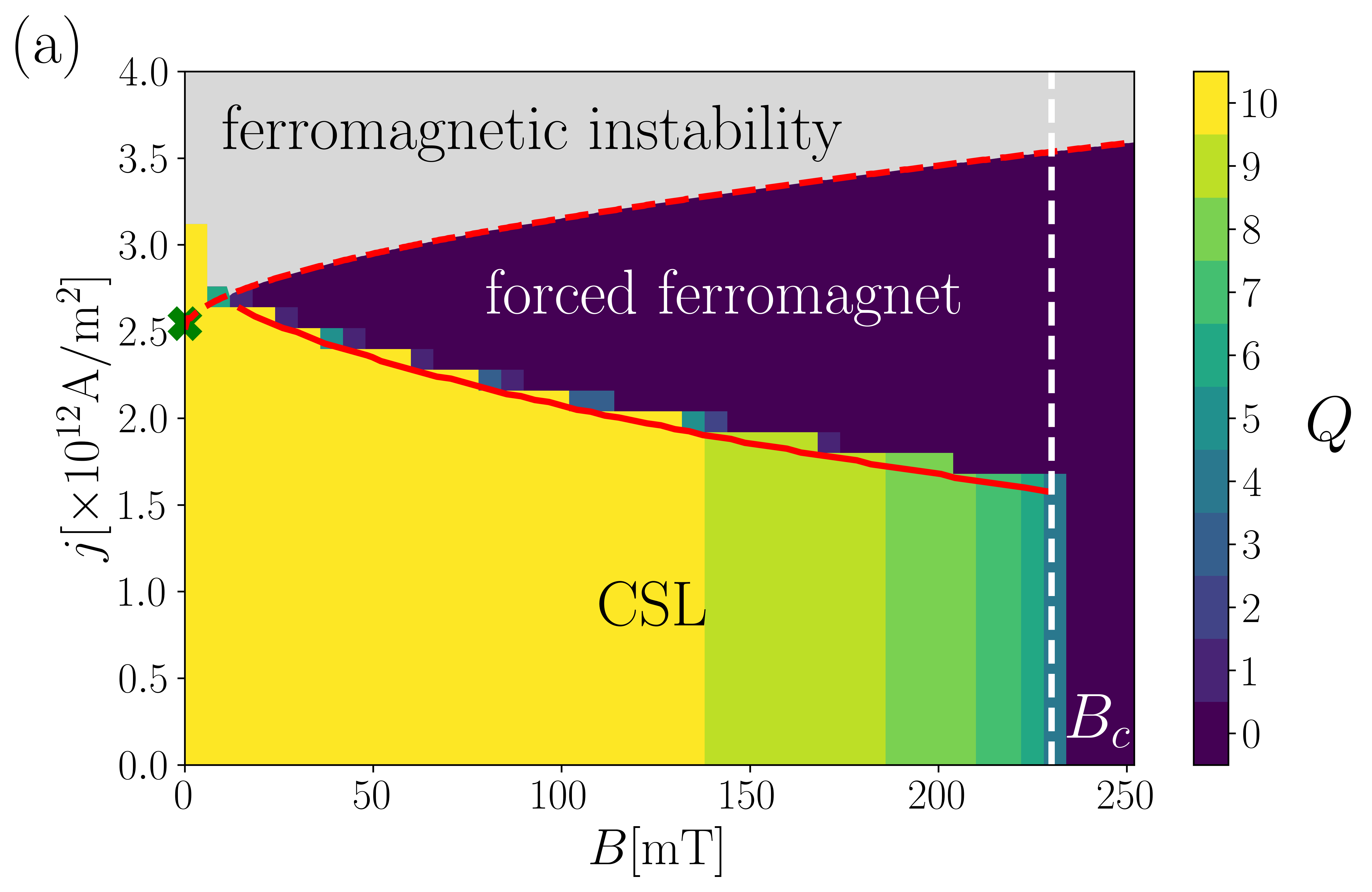}

\includegraphics[width=8.0cm]{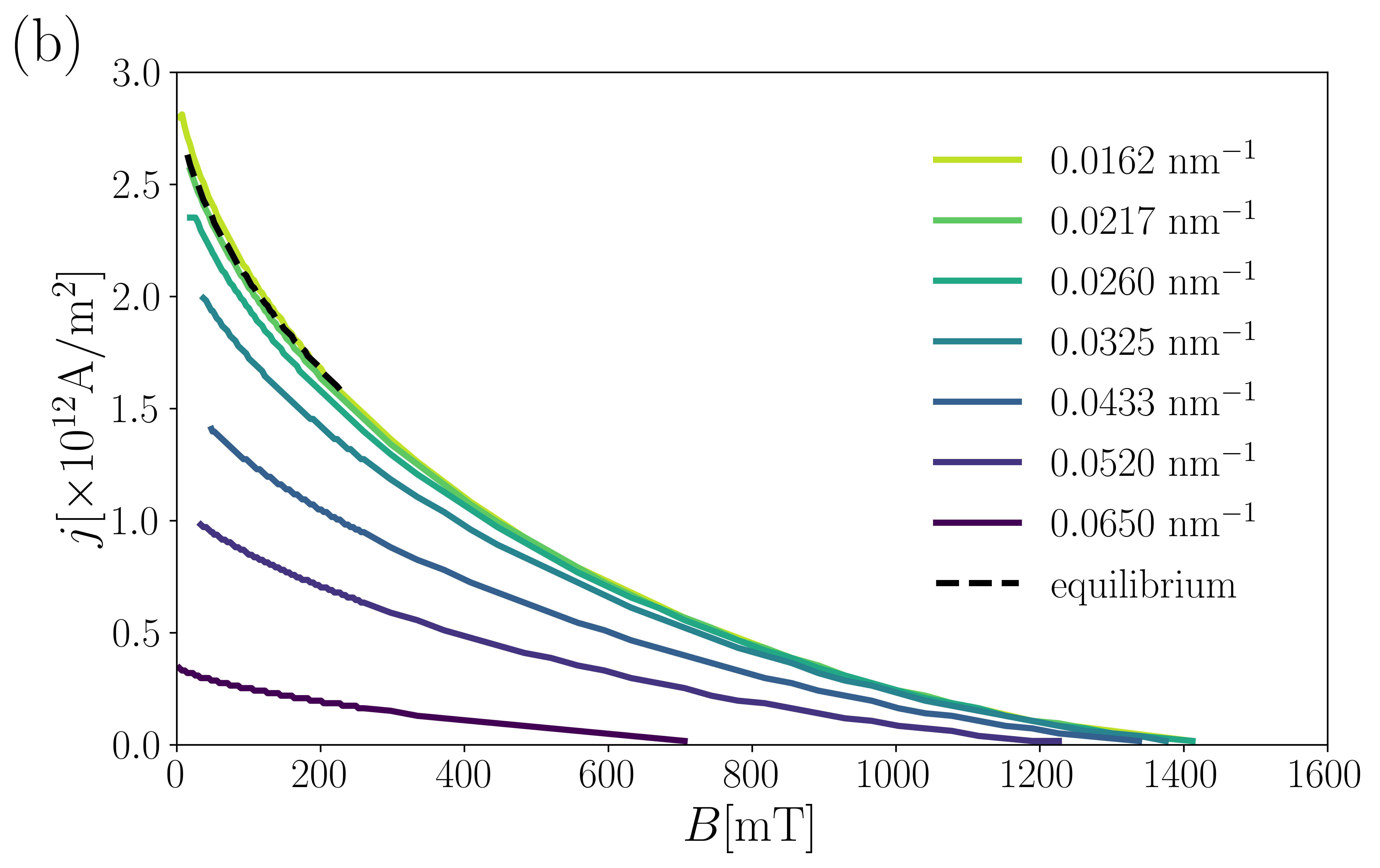}
\caption{
(a) The $j$-$B$ phase diagram for a monoaxial helimagnet. The color code indicates the value of the winding number $Q$ which, due to periodic boundary conditions, only takes integer values ($0\leq Q\leq 10$ for the equilibrium state in a system of size $R=500\un{nm}$) for the final magnetization state after a 50 ns long pulse of intensity $j$ at each value of the magnetic field $B$. The solid red line represents the analytic limit for the stability of the CSL. The dashed red line represents the analytic limit for the stability of the ferromagnetic state (which is unstable within the gray region). The dashed white line represents the critical field $B_{c}=230$ mT. The green cross represents the critical current for the helical state at $B=0\un{mT}$. Its value $j\approx2.51\times10^{12}\un{A/m^{2}}$ is very close to the value of the critical current for the stability of ferromagnetic state ($j\approx2.54\times10^{12}\un{A/m^{2}}$).
(b) The stability limit of the CSL at constant density of solitons, as indicated in the key. The dashed black line represents the stability limit for the equilibrium state (shown in (a)), in which the density of chiral solitons varies with the magnetic field.
\label{fig:phase_diagram}}
\end{figure}

\section{\label{sec:phase_diagram}Phase diagram}

Extending the previous analysis to different values of $j$ and $B$ it is possible to construct the phase diagram shown in Fig. \ref{fig:phase_diagram}(a). From micromagnetic simulations the winding number $Q$ in the final state after a $50 \un{ns}$ pulse of current is obtained. The winding number is computed as $Q=\sum_{i} \text{arcsin}\left[(\bm{\hat{n}}_{\perp,i}\times\bm{\hat{n}}_{\perp,i+1})\cdot\bm{\hat{z}}\right]$, 
where the sum runs over the number of cells along the chiral axis, $\bm{\hat{n}}_{\perp,i}=\bm{n}_{\perp,i}/\lvert\bm{n}_{\perp,i}\rvert$ and $\bm{\hat{z}}\cdot\bm{n}_{\perp,i}=0$, and counts the number of chiral solitons winding around the chiral axis in the system. It is important to note that this definition of $Q$ does not involve the evaluation of derivatives (through finite differences). This implies that the value of $Q$ is well quantized, taking integer values, and its value does not depend on the mesh size used in the discretization of the system. The computation of $Q$, as introduced here, resembles the method for the computation of the topological charge (or skyrmion number) in two dimensional systems using a lattice-based approach \cite{kim2020quantifying}.

The region with a gradient scale of colors from yellow to dark blue corresponds to $j<j_c$ where we find a CSL with the number of CSs decreasing from $N=10$ to $N=0$ for increasing magnetic fields. The region in dark blue corresponds to $Q=0$, and this means that the magnetization texture is not winding around the chiral axis, which eventually result in a ferromagnetic state. For $j=0$ we observe the typical behavior of a monoaxial chiral magnet in a transverse magnetic field. Since in our simulations we consider a system of size $R=500 \un{nm}$, and at zero magnetic field the period of the magnetic texture is $L_{0}\approx 48 \un{nm}$, the number of chiral solitons is thus $Q=10$. This value decreases down to $Q=0$ as the magnetic field grows and the system reaches the ferromagnetic state at $B_c$.
The solid red line corresponding to $j_c=j_{c}(B)$ was obtained using the stability analysis and agrees with the results from micromagnetic simulations. 
It is observed that the winding number does not change with the current except at the transition point, where it drops to zero discontinuously. A change in $Q$ involves the removal of a chiral soliton and this could occur in two ways, either through the edges of the system or destroying locally a chiral soliton. Since we simulate infinite systems, through the implementation of periodic boundary conditions, the first mechanism is forbidden due to the absence of edges. 
Since $Q$ is conserved when the current is increased below $j_c(B)$, the local destruction of CSs is not observed in our numerical simulations, presumably due to the topological protection of the CSL state. However an unwinding process of individual CSs could be present at low magnetic fields \cite{masell2020manipulating}.

The instability of the ferromagnetic state occurs for $j>j_{c}^{FM}(B)$ due to the current-assisted excitation of spin waves and is a well-known fact, usually encountered in different models of ferromagnets~\cite{Bazaliy98, Fernandez04, Tserkovnyak06, Masell2020}. In Fig. \ref{fig:phase_diagram}(a) the ferromagnetic state is unstable in the gray region and the critical current $j_{c}^{FM}(B)$ is represented by the dashed red line.

Above $j_{c}^{FM}$ the magnetization field does exhibit neither spatial nor temporal structure. It is important to note that for $B\lesssim 12 \un{mT}$ the CSL is driven directly to the region where the ferromagnet is unstable, without passing through a ferromagnetic state.
The value at $B=0\un{mT}$ can be directly computed to obtain $j_{c}(0)\approx2.51\times10^{12} \un{A/m^{2}}$ (green cross in Fig. \ref{fig:phase_diagram}(a)). In this region the necessary computation time to reach $j_{c}$ using micromagnetic simulations increases noticeable. Since we used a maximum time of $50\un{ns}$, the stability limit shown in Fig.~\ref{fig:phase_diagram}(a) is slightly larger than the analytical limit for $j_{c}(B)$ when $B \to 0$. Within this region, random fluctuations could also lead to an unwinding dynamical process, gradually reducing the number of CSs \cite{masell2020manipulating}.

The phase diagram shown in Fig. \ref{fig:phase_diagram}(a) corresponds to the equilibrium state, in which the density of chiral solitons minimizes the energy (at zero current), and thus varies with the magnetic field. 
However, due to the protection of the topologically non trivial states, each metastable states characterized by the density of solitons has its own critical current, $j_c(B)$, which is displayed in Fig. \ref{fig:phase_diagram}(b) for different values of the density of solitons. For comparison, the critical current corresponding to the equilibrium state is also shown (dashed black line). We see that $j_{c}(B)$ decreases both with $B$ and with the density of solitons.

\section{\label{sec:conclusions}Discussion and Conclusions}

We have described how the CSL responds to an applied current beyond the weak current density and weak magnetic field regimes ($B$ small compared to $B_c$). For each value of the magnetic field we find a critical current $j_{c}$ depending on the density of solitons. In the subcritical regime ($j<j_{c}$) the velocity-current response is linear and does not depend on the density of solitons. The steady finite velocity regime is accompanied by a conical distortion of the CSL, similar to the one observed when applying magnetic fields with a finite $z$ component. The magnitude of the applied current governs two properties of the conical distortion: the cross section of the cone decreases with the current, while the deviation of the cone axis, with respect to the chiral axis, increases with the current.

In the supercritical regime, $j>j_{c}$, the CSL is destabilized and the system reaches a ferromagnetic state with the magnetization oriented along the external field (except within the range $0\un{mT}\leq B \lesssim 12\un{mT}$). Even in this supercritical regime the evolution of the CSL to the ferromagnetic state can still be described, qualitatively, by an oriented conical deformation, but with strong deviations. 

The velocity of the CSL dragged by a spin polarized current has been already studied in Ref.~\onlinecite{Tokushuku17}, assuming weak magnetic fields. In that article the authors find that the terminal velocity for the CSL exhibits a weak dependence on the magnetic field for $B\ll B_{c}$, that can be recast as an approximately constant velocity, in agreement with our findings.
In addition, in the calculations of Refs. \onlinecite{Kishine10} and \onlinecite{Tokushuku17} the authors considered $\theta\approx\pi/2$. We go beyond this limit by considering that the spin polarized current can induce pronounced distortions in the structure of the CSL in which $\theta(z)$ is allowed to significantly depart from $\theta(z)=\pi/2$. 

Let us end the article with a brief discussion about the practical relevance of the results reported in this work. Firstly, the stability limit of the CSL imposes a constraint on the velocity of the CSL. That is, at a given magnetic field, $v$ can not exceed the critical velocity $v_{c}(B)=\frac{\beta b_{j}}{\alpha}j_{c}(B)$. Since $j_{c}(B)$ is a decreasing function of $B$, $v_{c}(B)\leq v_{c}(0)$, and that in turn implies for CrNb$_3$S$_6$ that the maximum velocity for a CSL is $v=v_{c}(0)\approx2600\un{m/s}$ (for $\alpha=0.01$ and $\beta=0.02$). Finally, although not shown in detail here, it is important to mention that once the ferromagnetic state is destabilized, and after turning off the current, the system evolves to a CSL with a variable number of CSs. Since the forced ferromagnet and CSL have very different magnetoresistive responses~\cite{Togawa13,Togawa15}, the dynamics described here allows a write/erase mechanism by using two currents $j_w$ and $j_e$ to switch between states with high and low magnetoresistance.
For instance, lets consider two current pulses of values $j_e$ and $j_w$ with $j_e < j_w$ and such that $j_{c}(B)<j_e<j_{c}^{FM}(B)$ and $j_w>j_{c}^{FM}(B)$. By applying a pulse of intensity $j_e$ to the CSL the system is driven into a ferromagnetic state which is then metastably retained when the current is removed, i.e a low-magnetoresistive state is retained. If we then apply a pulse with intensity $j_{w}$ the system goes beyond the ferromagnetic instability and then relaxes to a CSL,
which would correspond to a high-magnetoresistive state. After a sequence $j_e$-$j_w$ current pulses the initial and final CSL would, in general, have different number of CSs, which would comprise a small difference between high-magnetoresistive states but would not drastically affect the possible observation of two well resolved high- and low- magnetoresistive states. The results discussed here could therefore be relevant for the development of spintronic devices.

\begin{acknowledgments}
The authors acknowledge support by Grants No PGC-2018-099024-B-I00-ChiMag from the Ministry of Science and Innovation (MCIN) of Spain, SpINS-OTR2223 from CSIC/MCIN and DGA-M4 from the Diputación General de Aragón, Spain.
This work was also supported by the Grant No. PICT 2017-0906 from the Agencia Nacional de Promoción Científica y Tecnológica, Argentina.
\end{acknowledgments}

\appendix

\section{\label{sec:app-bvp} Solution of the boundary value problem for the steady state}

Let us discuss in this appendix some details about the boundary value problem which determines the steady states. It is set out by Eqs.~\eqref{eq:stationary1}, \eqref{eq:stationary2}, and \eqref{eq:BC1}, and therefore has to be solved in the interval $[-w_L,w_L]$.

The applied current density, $j$, and the steady state velocity, $v$, appear in the steady state equations  \eqref{eq:stationary1} and \eqref{eq:stationary2} through the combinations $\Omega$ and $\Gamma$, which may be seen as the natural parameters for the boundary value problem that determines the steady state. Notice that there is a one-to-one correspondence between the pairs $(j,v)$ and $(\Omega,\Gamma)$. To find the steady state we adopted the following strategy. For given values of $\Omega$, $\Gamma$, and $\theta_L$, we solve the boundary value problem given by Eqs.~\eqref{eq:stationary1} and \eqref{eq:stationary2} and the boundary conditions

\be
\begin{split}
 &\varphi(-w_L) = 0, \\
 &\varphi(w_L) = 2 \pi, \\
 &\theta(-w_L) = \theta_L, \\
 &\theta(w_L) = \theta_L. \\
\end{split}
\label{eq:BCaux}
\ee

\begin{figure}[t!]
\includegraphics[width=4.2cm]{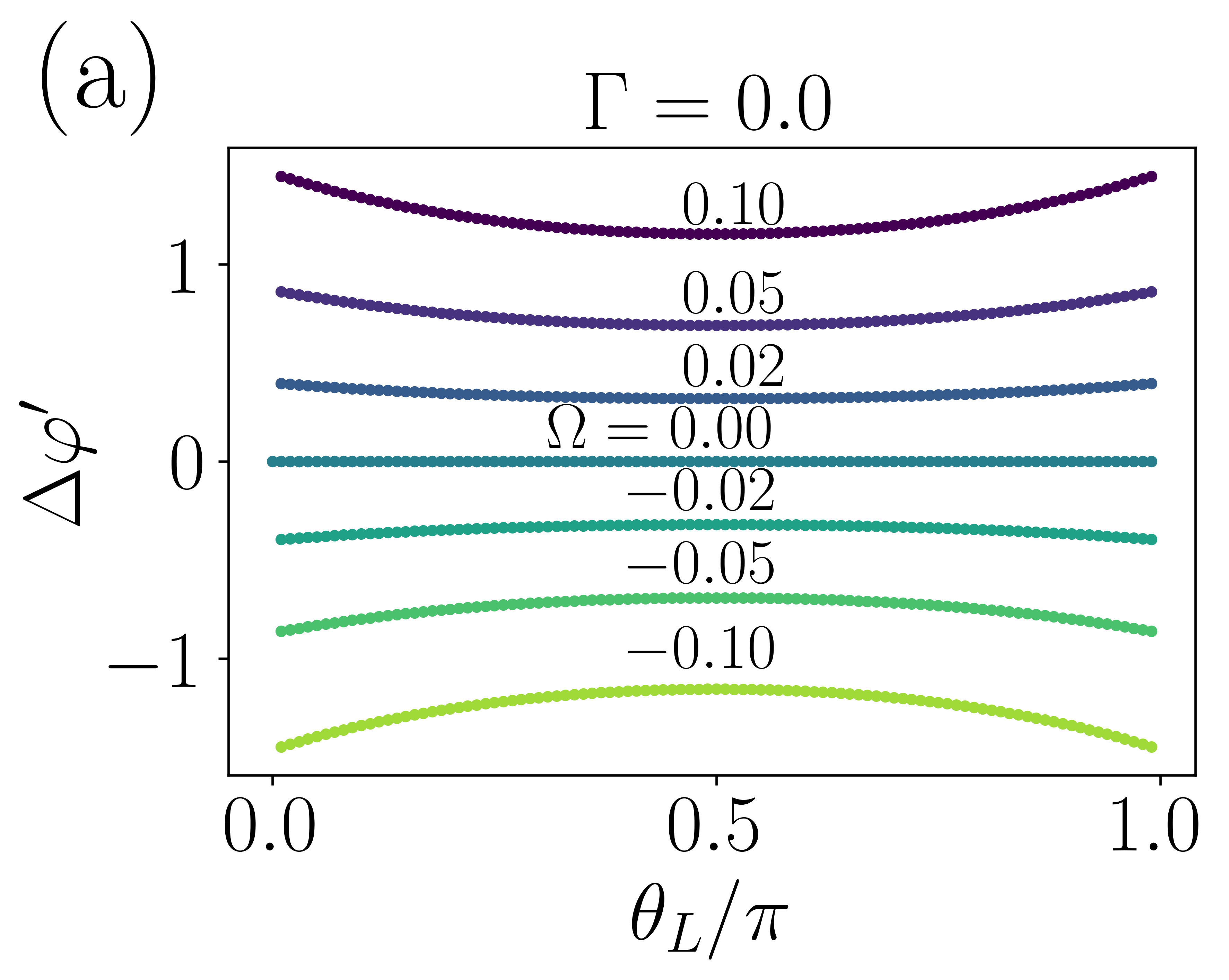}
\includegraphics[width=4.2cm]{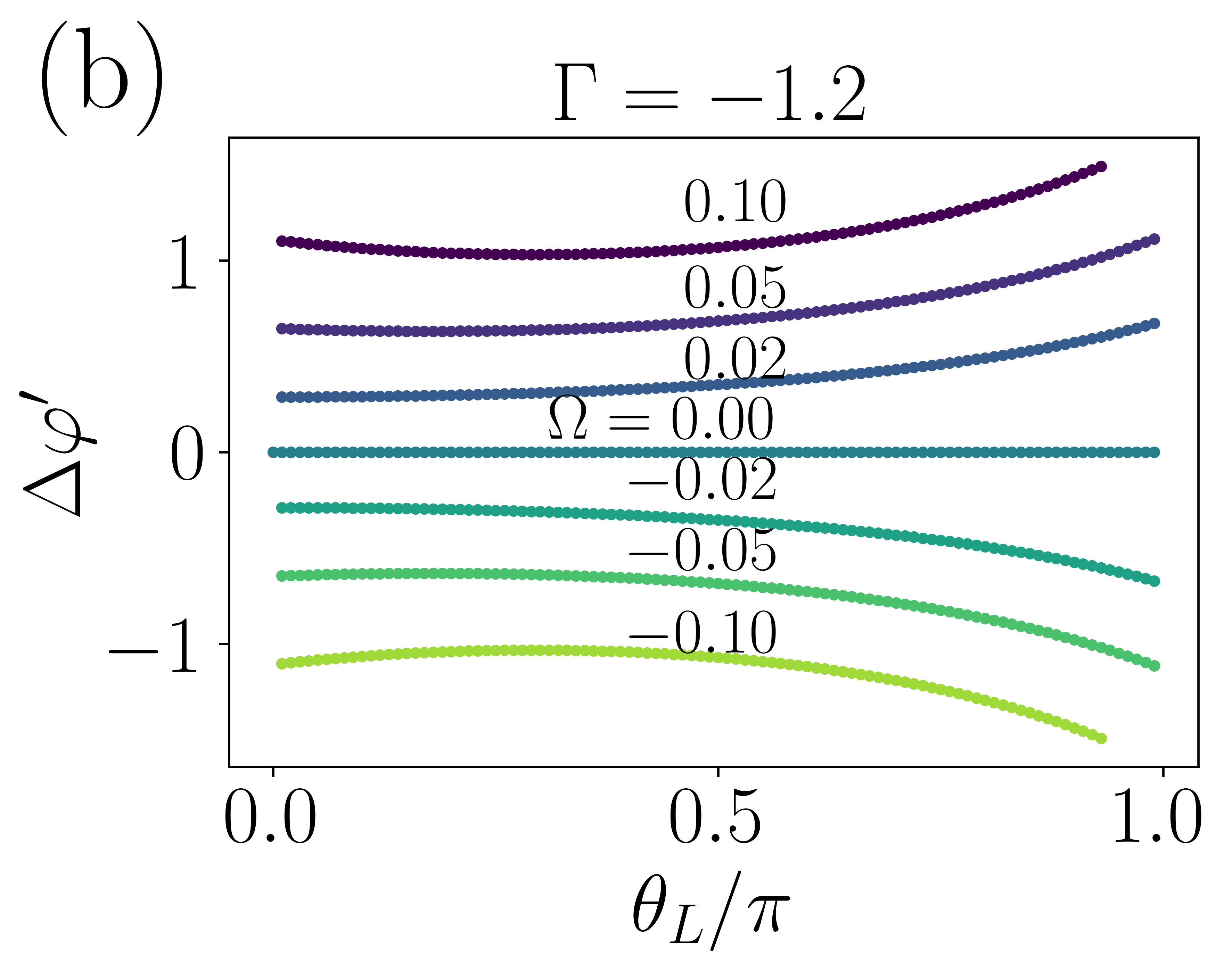}
\includegraphics[width=4.2cm]{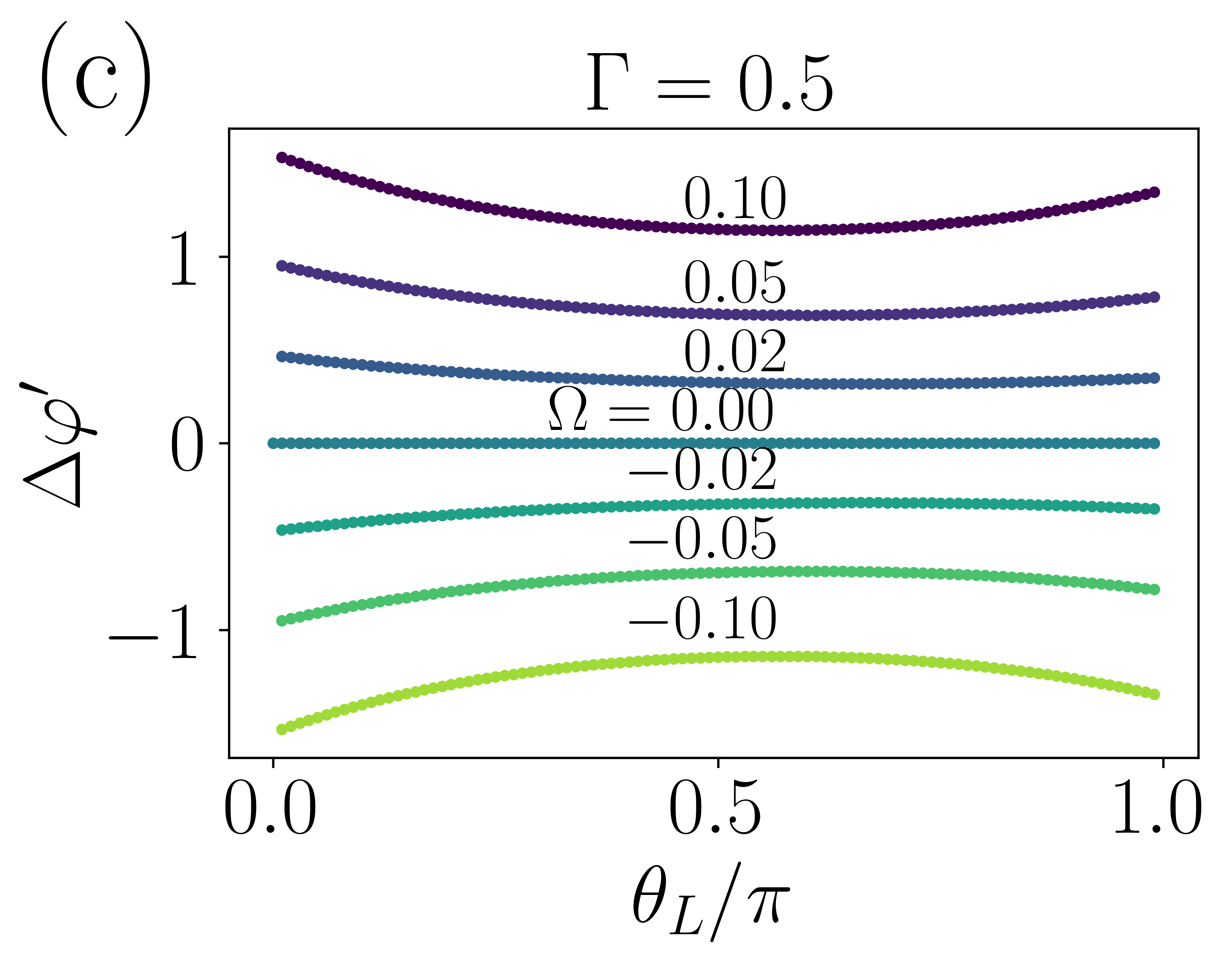}
\includegraphics[width=4.2cm]{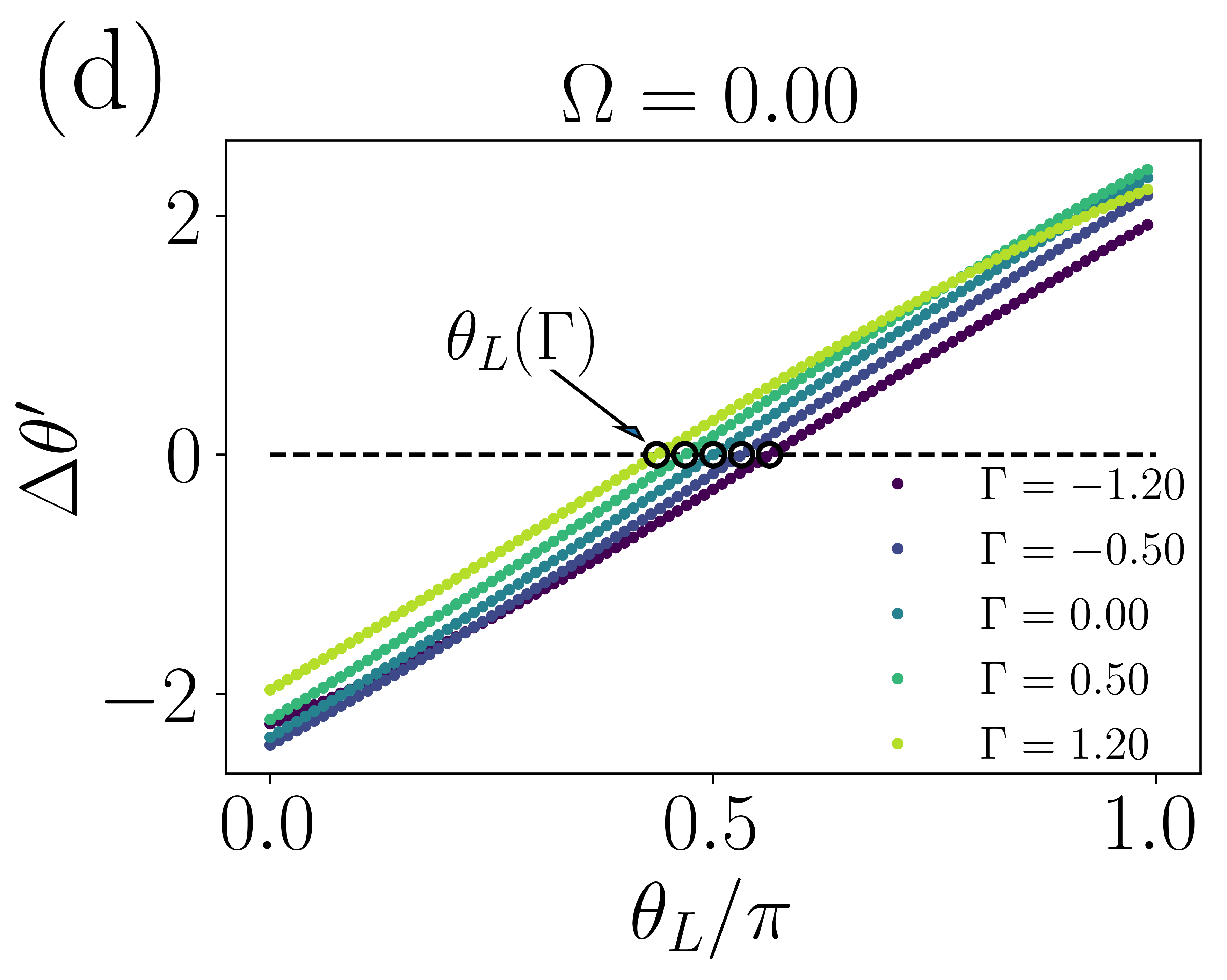}
\caption{
Dependence of $\Delta \varphi' = \varphi'(w_L) - \varphi'(-w_L)$ and $\Delta \theta' = \theta'(w_L) - \theta'(-w_L)$ on the boundary tilt angle $\theta_L$ for different values of $\Omega$ and $\Gamma$. In (a), (b) and (c) $\Delta \varphi'(\theta_L)$ is shown for different $\Omega$ values and fixed $\Gamma = 0$, $-1.2$, and $0.5$, respectively. $\Delta \varphi' = 0$ only when $\Omega = 0$, irrespective of the value of $\theta_L$.
(d) shows $\Delta \theta'(\theta_L)$ for $\Omega = 0$ and different $\Gamma$ values. The values of $\theta_L(\Gamma,\Omega=0)$ satisfying $\Delta \theta'=0$ and $\Delta \varphi' = 0$ are indicated as open-black circles. $\theta_L(\Gamma,\Omega=0)$ results in $\theta_L(j)$ shown in Fig.~\eqref{fig:critic-sol}.
\label{fig:numerical_solution}}
\end{figure}

In general, a solution to this problem can numerically be found for different values of $\Omega$, $\Gamma$, and $\theta_L$.
We solved this problem numerically using a finite difference method with centered finite differences for the derivatives. The resulting nonlinear equations were solved by a relaxation method.
A solution of the boundary value problem which we solve numerically, associated to the boundary conditions \eqref{eq:BCaux}, is a solution of the steady state boundary value problem, associated to the boundary conditions \eqref{eq:BC1} and \eqref{eq:BC2}, if and only if
\bea
 \Delta \varphi' &=& \varphi'(w_L) - \varphi'(-w_L) = 0, \\
 \Delta \theta' &=& \theta'(w_L) - \theta'(-w_L) = 0.
\eea
Clearly, these additional conditions will be fulfilled only at specific values of $\Omega$, $\Gamma$, and $\theta_L$. It turns out that $\Delta \varphi' = 0$ if and only if $\Omega=0$. Some examples are shown in Figs.~\ref{fig:numerical_solution}(a)-(c), where $\Delta \varphi'$ is plotted as a function of $\theta_L$ for different values of $\Gamma$ and $\Omega$. Therefore, we are forced to set $\Omega=0$, what implies the linear relation of $v$ and $j$ given by Eq.~\eqref{eq:v_vs_j}, and that $\Gamma$ is proportional to $j$.

With $\Omega=0$ and for a given value of $\Gamma$, condition  $\Delta \theta'=0$ is satisfied only for specific values of $\theta_L$. This is illustrated in Fig.~\ref{fig:numerical_solution}(d). We can use these values of $\theta_L$ to characterize the steady solutions at given $\Gamma$. They are displayed as a function of $j$ in Fig.~\ref{fig:critic-sol}. Finally, the boundary value problem associated to the boundary conditions \eqref{eq:BCaux} has no solution if $|\Gamma|$ is larger than a certain value $|\Gamma_c|$ which depends on the rest of the parameters of the model (the applied field, the anisotropy energy, etc.)
This means there is no steady motion state for $j>j_c$.

\section{\label{sec:app-stability} Stability analysis of the steady solution}

Let $\bm{n}_0$ be a steady state and consider a perturbation about it described by two fields $\xi_1$ and $\xi_2$ as in Eq. \eqref{eq:pert}. We choose 
\be
\bm{e}_1 = \partial \bm{n} / \partial\theta, \quad \bm{e}_2 = \bm{n}_0 \times \bm{e}_1,
\ee
where $\bm{n}$ is given by Eq. \eqref{eq:polar_param} and $\theta$ and $\varphi$ are the solution of the boundary value problem, defined by Eqs. \eqref{eq:stationary1}, \eqref{eq:stationary2}, and \eqref{eq:BC1}, which determines the steady state.
Remember that while $\xi_1$ and $\xi_2$ are functions of the three coordinates $x$, $y$, $z$, and of time, $t$, the vectors $\bm{n}_0$, $\bm{e}_1$ and $\bm{e}_2$ are functions of the single variable $w=q_0(z-vt)$. Hence, as discussed in Section \ref{subsec:steady_bvp}, it is convenient to perform a change of variable and consider $\xi_1$ and $\xi_2$ as functions of $t$, $x$, $y$, and $w$.

Defining the two-component column vector $\xi=(\xi_1,\xi_2)^T$ the dynamics of the perturbation is governed by the linearized LLG equation (Eq. \eqref{eq:lin_LLG}), with the linear operator $\mathcal{S}$ given by
\be
\begin{gathered}
  \mathcal{S} = \omega_0\left[\Big(J-\alpha I\Big) K
 + \frac{b_j j}{v_0} \frac{\beta-\alpha}{\alpha}\Big(I+\alpha J\Big) U\right],
\end{gathered}
\ee
where $\omega_0=v_0q_0/(1+\alpha^2)$, 
\be
I = \left(
 \begin{array}{cc}
  1 & 0 \\
  0 & 1
 \end{array}
 \right), \quad
 J = \left(
 \begin{array}{cc}
  0 & -1 \\
  1 & 0
 \end{array}
 \right),
\ee
$K$ is a $2\times 2$ matrix of operators with matrix elements 
\bea
 K_{11} &=& -q_0^{-2}\,\nabla_\perp^2 - \partial_w^2 + \cos 2 \theta\big(\varphi^{\prime\,2}-2\varphi'+\kappa\big)  \nonumber \\
          & & + h_y\sin\theta\cos\varphi , \\[4pt]
 K_{12} &=& 2(\varphi'-1)\cos \theta\,\partial_w + \cos \theta \varphi'', \\[4pt]
 K_{21} &=& -2(\varphi'-1)\big(\cos\theta\,\partial_w-\sin\theta\theta'\big)- \cos \theta \varphi'' \\[4pt] 
 K_{22} &=& -q_0^{-2}\,\nabla_\perp^2 - \partial_w^2 - \theta'^2  + \cos^2 \theta\big(\varphi^{\prime\,2}-2\varphi'+\kappa\big) \nonumber \\
             & & + h_y\sin\theta\cos\varphi,
\eea
with $\nabla_\perp^2=\partial_x^2+\partial_y^2$, and
\be
 U =
 \left(
  \begin{array}{cc}
    \partial_w & -\cos \theta \varphi' \\
    \cos \theta \varphi' & \partial_w
  \end{array}
 \right).
\ee
The primes stand for derivatives with respect to $w$.
The functions $\theta(w)$ and $\varphi(w)$ characterize the steady solution,
which is stable if the spectrum of the $\mathcal{S}$ operator lies on the left half plane of the complex plane, that is, if all of its
eigenvalues have non positive real part.

Since the functions $\theta$ and $\varphi$ are periodic, with the period of the CSL, $\mathcal{S}$ is a periodic operator (it commutes with the lattice translations). 
Therefore, we used the Bloch-Floquet theorem to reduce the spectral problem of $\mathcal{S}$ to the spectral problem of a related operator which acts on the space of periodic functions.
The eigenvalue of $\mathcal{S}$ with largest real part has been estimated by discretizing the operator acting on periodic functions and obtaining the relevant part of its spectrum with an Arnoldi method.

\section{The CSL velocity from autocorrelation}
\label{sec:app-autocorrelation}

In order to obtain the velocity of the CSL from the simulations, and considering the intrinsic periodicity of the system, we compute the autocorrelation function
\be
C(t)=\langle \bm{n}(z,0)\cdot\bm{n}(z,t)\rangle=\frac{1}{R}\int_{0}^{R}\bm{n}(z,0)\cdot\bm{n}(z,t)dz,
\ee
where $\bm{n}(z,t)$ is the magnetization field at time $t$ and position $z$. If the dynamical evolution of the magnetization field corresponds to a steady and rigid translation of the CSL, then $C(t)$ presents a periodic structure characterized predominantly by a single frequency.
For a CSL of period $L$, we can expand each component of the magnetization in the form
\be
n_{i}(z)=\sum_{k}A_{i,k}\cos \left(z \frac{k 2\pi}{L}+\phi_{k} \right),
\ee
where $i=x,y,z$.
Since we fix the number $N$ of chiral solitons in the system of size $R$ we have that $L=R/N$.
Then we find:
\be
\int_{0}^{R}n_{i}(z)n_{i}(z-z_{0})dz=\frac{R}{2}\sum_{k}\left\lbrace A_{i,k}^{2}\cos \left(\frac{k z_{0}2\pi N}{R} \right)\right\rbrace.
\ee
If we replace $z_{0}=v\,t$ and sum over $i=x,y,z$ we get 
\be
\label{eq:fourier_c}
C(t)=\frac{1}{2}\sum_{k}\left\lbrace\left[\sum_{i=x,y,z}A_{i,k}^{2}\right]\cos\left(\frac{N 2 \pi k v t}{R}\right)\right\rbrace,
\ee
which represents the Fourier expansion of the $C(t)$ function in terms of the frequencies $\nu_{k}=2 \pi k v N/R$.
In practice, it results that $A_{k,i}\approx 0$ for $|k|>1$, and the Fourier expansion in Eq.~\eqref{eq:fourier_c} is dominated essentially by the $\nu_{1}$ term.
The velocity of the CSL can finally be obtained from the lowest non-zero frequency $\nu_{1}$, $v=\frac{\nu_{1}R}{2\pi N}$.

%\bibliography{references}% Produces the bibliography via BibTeX.

\end{document}